\documentclass[]{aa}
\usepackage{graphicx}
\usepackage{txfonts}
\usepackage{hyperref}
\usepackage{float}
\usepackage{amsmath,amsfonts,amssymb}
\usepackage{natbib}
\bibpunct{(}{)}{;}{a}{}{,}

\newcommand*\justify{%
  \fontdimen2\font=0.4em
  \fontdimen3\font=0.2em
  \fontdimen4\font=0.1em
  \fontdimen7\font=0.1em
  \hyphenchar\font=`\-
}
\newcommand{\code}[1]{\justify\texttt{#1}}

\newcommand{\parallax}{\varpi}

\newcommand{\LAMOST} {{LAMOST}}
\newcommand{\SDSS} {{SDSS}}
\newcommand{\DESI}{{DESI}}
\newcommand{\Gaia}{{\it Gaia}}

\newcommand{\Catwise}{{CatWISE2020}}
\newcommand{\bannjos}{\texttt{BANNJOS}}

\begin{document} 

\title{J-PLUS: Bayesian object classification with a strum of \bannjos{}}

\author{A.~del Pino\inst{\ref{CEFCA}}
\and C.~L\'opez-Sanjuan\inst{\ref{CEFCA}}
\and A. Hern\'an-Caballero\inst{\ref{CEFCA}}
\and H. Dom\'{\i}nguez-S\'anchez\inst{\ref{CEFCA}}
\and R. von Marttens\inst{\ref{UFB},\ref{PPGCosmo}}
\and J. A. Fern\'andez-Ontiveros\inst{\ref{CEFCA}}
\and P. R. T. Coelho\inst{\ref{USP}}
\and A. Lumbreras-Calle\inst{\ref{CEFCA}}
\and J. Vega-Ferrero\inst{\ref{CEFCA}}
\and F. Jimenez-Esteban\inst{\ref{INTA}}
\and P. Cruz\inst{\ref{INTA}}
\and V. Marra\inst{\ref{DFUFES}, \ref{INAF}, \ref{IFPU}}
\and M. Quartin\inst{\ref{IFRio},\ref{OValongo},\ref{PPGCosmo}}
\and C. A. Galarza\inst{\ref{ON}}
\and R.~E.~Angulo\inst{\ref{DIPC},\ref{ikerbasque}}
\and A.~J.~Cenarro\inst{\ref{CEFCA}}
\and D.~Crist\'obal-Hornillos\inst{\ref{CEFCA2}}
\and R.~A.~Dupke\inst{\ref{ON},\ref{MU},\ref{Alabama}}
\and A.~Ederoclite\inst{\ref{CEFCA}}
\and C.~Hern\'andez-Monteagudo\inst{\ref{IAC},\ref{ULL}}
\and A.~Mar\'{\i}n-Franch\inst{\ref{CEFCA}}
\and M.~Moles\inst{\ref{CEFCA2}}
\and L.~Sodr\'e Jr.\inst{\ref{USP}}
\and J.~Varela\inst{\ref{CEFCA2}}
\and H.~V\'azquez Rami\'o\inst{\ref{CEFCA}}
}

\institute{
    Centro de Estudios de F\'{\i}sica del Cosmos de Arag\'on (CEFCA), Unidad Asociada al CSIC, Plaza San Juan 1, 44001 Teruel, Spain\\\email{andresdelpinomolina@gmail.com}\label{CEFCA}
    \and
    Instituto de F\'{\i}sica, Universidade Federal da Bahia, 40170-155, Salvador, BA, Brazil\label{UFB}
    \and
    PPGCosmo, Universidade Federal do Esp\'{\i}rito Santo, 29075-910, Vit\'oria, ES, Brazil\label{PPGCosmo}
    \and
    Instituto de Astronomia, Geof\'{\i}sica e Ci\^encias Atmosf\'ericas, Universidade de S\~ao Paulo, 05508-090 S\~ao Paulo, Brazil\label{USP}
    \and
    Centro de Astrobiolog\'{\i}a, CSIC-INTA, Camino bajo del castillo s/n, E-28692, Villanueva de la Can\~ada, Madrid, Spain\label{INTA}
    \and
    Departamento de F\'{\i}sica, Universidade Federal do Esp\'{\i}rito Santo, 29075-910, Vit\'oria, ES, Brazil\label{DFUFES}
    \and
    INAF, Osservatorio Astronomico di Trieste, via Tiepolo 11, 34131 Trieste, Italy\label{INAF}
    \and
    IFPU, Institute for Fundamental Physics of the Universe, via Beirut 2, 34151, Trieste, Italy\label{IFPU}
    \and
    Instituto de F\'{\i}sica, Universidade Federal do Rio de Janeiro, 21941-972, Rio de Janeiro, RJ, Brazil\label{IFRio}
    \and
    Observat\'orio do Valongo, Universidade Federal do Rio de Janeiro, 20080-090, Rio de Janeiro, RJ, Brazil\label{OValongo}
    \and
    Observat\'orio Nacional - MCTI (ON), Rua Gal. Jos\'e Cristino 77, S\~ao Crist\'ov\~ao, 20921-400 Rio de Janeiro, Brazil\label{ON}
    \and
    Donostia International Physics Centre (DIPC), Paseo Manuel de Lardizabal 4, 20018 Donostia-San Sebastián, Spain\label{DIPC}
    \and
    IKERBASQUE, Basque Foundation for Science, 48013, Bilbao, Spain\label{ikerbasque}
    \and
    University of Michigan, Department of Astronomy, 1085 South University Ave., Ann Arbor, MI 48109, USA\label{MU}
    \and
    University of Alabama, Department of Physics and Astronomy, Gallalee Hall, Tuscaloosa, AL 35401, USA\label{Alabama}
    \and
    Instituto de Astrof\'{\i}sica de Canarias, La Laguna, 38205, Tenerife, Spain\label{IAC}
    \and
    Departamento de Astrof\'{\i}sica, Universidad de La Laguna, 38206, Tenerife, Spain\label{ULL}
    \and
    Centro de Estudios de F\'{\i}sica del Cosmos de Arag\'on (CEFCA), Plaza San Juan 1, 44001 Teruel, Spain\label{CEFCA2}
}

\date{Received XX XX 2024; accepted XX XX XX}

\abstract
{With its 12 optical filters, the Javalambre-Photometric Local Universe Survey (J-PLUS) provides an unprecedented multicolor view of the local Universe. The third data release (DR3) covers 3,192 deg$^2$ and contains 47.4 million objects. However, the classification algorithms currently implemented in the J-PLUS pipeline are deterministic and based solely on the sources morphology.
}
{
Our goal is to classify the sources identified in the J-PLUS DR3 images into stars, quasi-stellar objects (QSOs), and galaxies. For this task, we present \bannjos{}, a machine learning pipeline that utilizes Bayesian neural networks to provide the full probability distribution function (PDF) of the classification.
}
{
\bannjos{} is trained on photometric, astrometric, and morphological data from J-PLUS DR3, \Gaia{} DR3, and \Catwise{}, using over 1.2 million objects with spectroscopic classification from \SDSS{} DR18, \LAMOST{} DR9, \DESI{} Early Data Release, and \Gaia{} DR3. Results are validated on a test set of about $1.4 \times 10^5$ objects and cross-checked against theoretical model predictions.
}
{
\bannjos{} outperforms all previous classifiers in terms of accuracy, precision, and completeness across the entire magnitude range. It delivers over $95\%$ accuracy for objects brighter than $r = 21.5$ mag, and $\sim 90\%$ accuracy for those up to $r = 22$ mag, where J-PLUS completeness is $\lesssim 25\%$. \bannjos{} is also the first object classifier to provide the full probability distribution function (PDF) of the classification, enabling precise object selection for high purity or completeness, and for identifying objects with complex features, like active galactic nuclei with resolved host galaxies.
}
{
\bannjos{} has effectively classified J-PLUS sources into around 20 million galaxies, 1 million QSOs, and 26 million stars, with full PDFs for each, which allow for later refinement of the sample. The upcoming J-PAS survey, with its 56 color bands, will further enhance \bannjos{}'s ability to detail each source's nature.
}


\titlerunning{J-PLUS. Object calibration with \bannjos{}}

\authorrunning{del Pino et al.}

\maketitle

\section{Introduction}

Public photometric large digital sky surveys are revolutionizing our view of the Universe. Covering large areas of the sky ($\gtrsim 5000\,\text{deg}^2$), surveys such as POSS-II (Second Palomar Observatory Sky Survey, three optical $gri$ broad bands; \citealt{poss2}), SDSS (Sloan Digital Sky Survey, five optical $ugriz$ broad bands; \citealt{sdssdr7}), and VHS (VISTA Hemisphere Survey, three near-infrared $HJK_{\mathrm{s}}$ bands; \citealt{vhs}) have provided crucial information about the large-scale structures that dominate the Universe. This allows astronomers to better understand the nature of celestial objects and to study a wide range of phenomena, from star formation to the evolution of galaxy groups. The next generation of surveys, such as DES (Dark Energy Survey, five optical $ugriz$ broad bands; \citealt{des}), UHS (UKIRT Hemisphere Survey, two near-infrared $JK_{\mathrm{s}}$ bands; \citealt{uhs}), {\it Euclid} (three near-infrared $YJH$ broad bands; \citealt{euclid}), and LSST (Large Synoptic Survey Telescope, six optical $ugrizY$ broad bands; \citealt{lsst}), will push the current sensitivity limits and open the door to new fields such as time-domain astronomy.

Among the recent additions to the family of public photometric surveys is the Javalambre Photometric Local Universe Survey\footnote{\href{www.j-plus.es}{www.j-plus.es}} (J-PLUS; \citealt{jplus}), carried out at the Observatorio Astrof\'{i}sico de Javalambre (OAJ, Teruel, Spain; \citealt{oaj}) using the 83\,cm Javalambre Auxiliary Survey Telescope (JAST80) and T80Cam, a panoramic camera of 9.2k $\times$ 9.2k pixels that provides a $2\deg^2$ field of view (FoV) with a pixel scale of $0.55$ arsec pix$^{-1}$ \citep{t80cam}. J-PLUS aims to cover $8500\,\text{deg}^2$ of the northern sky hemisphere with an unprecedented set of 12 filters: five broad $ugriz$ bands plus seven narrow optical bands (refer to Table~\ref{tab:JPLUS_filters}). The vast dataset produced by J-PLUS has broad astrophysical applications, enhancing our understanding of the Universe. The third data release (DR3) spans 3192 deg$^2$ (2881 deg$^2$ after masking) and catalogs 47.4 million objects with improved photometric calibration \citep{Lopez-SanJuan2023}, offering an unprecedented multicolor view of the local Universe.\footnote{\href{https://www.j-plus.es/datareleases/data_release_dr3}{www.j-plus.es/datareleases/data\_release\_dr3}}

\begin{table} 
\caption{J-PLUS photometric system and limiting magnitudes.}
\label{tab:JPLUS_filters}
\centering 
        \begin{tabular}{l c c c}
        \hline\hline\rule{0pt}{3ex} 
        Filter $(\mathcal{X})$   & Central wavelength    & FWHM  & $m_{\rm lim}^{\rm DR3}$ \\\rule{0pt}{2ex} 
                &   [nm]                & [nm]           &  [mag]\tablefootmark{a}        \\
        \hline\rule{0pt}{2ex}
        $u$             &348.5  &50.8           &       20.8    \\ 
        $J0378$         &378.5  &16.8           &       20.8    \\ 
        $J0395$         &395.0  &10.0           &       20.8    \\ 
        $J0410$         &410.0  &20.0           &       21.0    \\ 
        $J0430$         &430.0  &20.0           &       21.0    \\ 
        $g$             &480.3  &140.9          &       21.8    \\ 
        $J0515$         &515.0  &20.0           &       21.0    \\ 
        $r$             &625.4  &138.8          &       21.8    \\ 
        $J0660$         &660.0  &13.8           &       21.0    \\ 
        $i$             &766.8  &153.5          &       21.3    \\ 
        $J0861$         &861.0  &40.0           &       20.4    \\ 
        $z$             &911.4  &140.9          &       20.5    \\ 
        \hline 
\end{tabular}
\tablefoot{
\tablefoottext{a} {Limiting magnitude (5$\sigma$, 3 arcsec diameter aperture) of J-PLUS DR3 \citep{Lopez-SanJuan2023}.}
}
\end{table}

Due to its photometric, flux-limited nature, J-PLUS images all astronomical sources down to its limiting magnitude without pre-selection. Identifying and classifying the observed objects within its footprint, such as stars and galaxies, is crucial. As with any photometric survey, this is one of the first steps in creating science-ready data products from J-PLUS data. Reliable object classification enables the study of specific astronomical sources and aids in the discovery of uncommon or new types of objects. Classification algorithms typically employ two complementary approaches: color-based and morphological. Color classifiers leverage the distinct positions of stars, galaxies, and quasi-stellar objects (QSOs) in color-color diagrams \citep[e.g.,][]{huang97,elston06,baldry10,saglia12,malek13}, while morphological classifiers distinguish between point-like and extended sources based on isophotal concentration \citep[e.g.,][]{kron80,reid96,odewahn04,vasconcellos11}. J-PLUS utilizes the \texttt{CLASS\_STAR} morphological classifier from the \texttt{SExtractor} photometry package \citep{sextractor}. However, this algorithm, which considers only object elongation, extension, and peak brightness, simplistically categorizes objects as "star" or "not star." Such a method is prone to misclassification, particularly when distinguishing between compact sources like stars, distant active galactic nuclei, or compact galaxies. Refined classification through manual inspection is possible but requires significant time and resources.

Some of these problems can be eased by incorporating prior information within a Bayesian framework \citep[e.g.,][]{sebok79,scranton02,henrion11,molino13}. In \citet{lopezsanjuan19}, the authors introduced the \code{sglc\_prob\_star} classification, which imposes priors based on concentration, broad-band colors, object counts, and distance to the Galactic plane to achieve more reliable results, especially for sources with low signal-to-noise (S/N) ratios. However, this method does not utilize the valuable multi-filter color information available in J-PLUS, and its output remains bimodal, distinguishing only between "compact" and "extended" sources. To achieve a reliable classification based on an object's nature, not just its morphology, it is essential to utilize all photometric information in J-PLUS. Prior works have attempted this using various classification methods; for instance, \citet{wang22} employed the 12 photometric bands to include the QSO class. Yet, their method was limited to high S/N sources, classifying only about 3.5 million sources out of the $\sim$13 million in J-PLUS's first data release. 

To classify the entire catalog, methods capable of handling missing data are necessary. Techniques based on machine learning, successfully implemented in other surveys, offer classification into two \citep[e.g.,][]{ball06sg, miller17} or three categories \citep[e.g.,][]{malek13}. In the particular case of J-PLUS, in \citet{vonmarttens24} authors applied eXtreme Gradient Boosting (\code{XGBoost}, \citealt{xgboost}) to classify objects into stars, galaxies and QSOs, being the first classifier to approach a three-class classification in this survey. Unlike Bayesian approaches, these methods provide a deterministic classification for each object. In the best of the cases, classifiers would offer deterministic probabilities for an object's class membership (e.g., star, QSO, galaxy) without uncertainty or correlation between the classes probabilities. However, models and training data carry inherent uncertainties, and the distinctions between classes are sometimes not clear, as in the case of partially resolved active galaxies blurring the lines between QSOs and galaxies. Reliable classification across the entire catalog requires models that accommodate these uncertainties, leveraging morphological, photometric, and external information to provide confidence intervals and correlations in their predictions. The inclusion of uncertainty and degeneration among classes is also critical in order to control the purity and completeness of object selections as well as biases in samples containing objects of mixed characteristics.

In this paper, we introduce \bannjos{}, a pipeline utilizing Bayesian Artificial Neural Networks (BANN) to classify J-PLUS objects into three categories: stars, QSOs, and galaxies. We evaluate the classification's quality and suggest methods to enhance the purity of the selected object samples. The paper is structured as follows: Section~\ref{sec:bannjos} outlines the \bannjos{} pipeline; Section~\ref{sec:bannjos_on_JPLUS} discusses its application to J-PLUS, highlighting specific execution details; Section~\ref{sec:Validation_model} presents classification results and model validation; Section~\ref{sec:Refining_selection} present examples of selection criteria; Section~\ref{sec:Comparison_previous} compares our method with three existing classifications for J-PLUS; Section~\ref{sec:Results_final_validation} present the results for the entire J-PLUS catalog and extra statistical validation tests; Section~\ref{sec:Caveats} explain known caveats, and finally, Section~\ref{sec:Conclusions} summarizes our findings and conclusions.

\section{\bannjos}\label{sec:bannjos}

\begin{figure}
\begin{center}
\includegraphics[trim={0cm 0cm 0cm 0},clip, width=\linewidth]{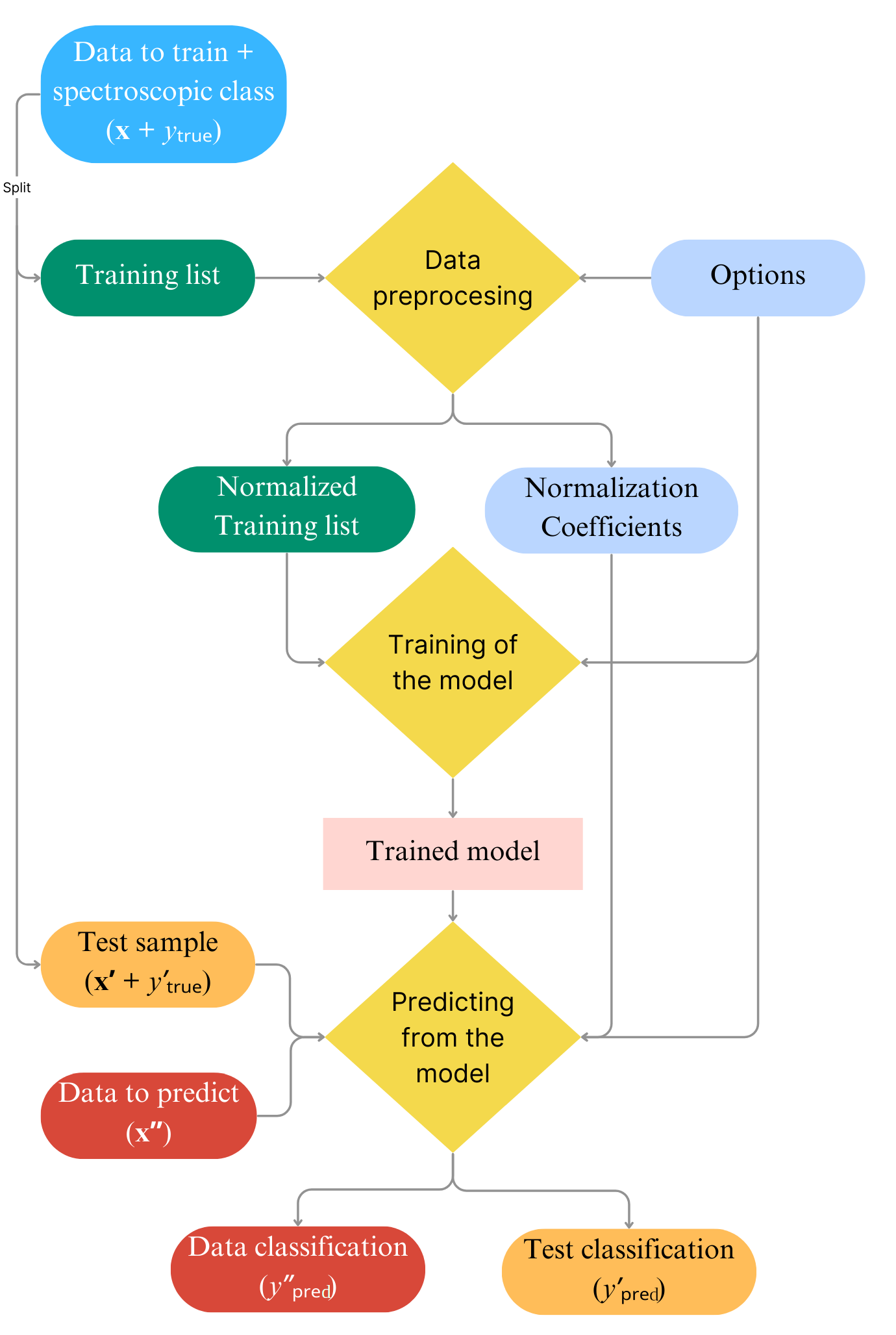}
\caption{Flowchart illustrating the flow of processing of \bannjos{}. Central components are represented by yellow rhombuses, while inputs and outputs are indicated by rounded rectangles. The procedure commences with the training data (depicted in blue, located at the top-left) and the options (illustrated in light blue, positioned at the top-right), which govern the behavior of \bannjos{}. Throughout the chart, colors show the data type: light blue for options or variables controlling the process, green for training data, orange for the test sample, and red for the data on which predictions are intended to be made.}
\label{fig:Bannjos_workflow}
\end{center}
\end{figure}

The Bayesian Artificial Neural Networks for the Javalambre Observatory Surveys, or \bannjos{}, is a publicly available\footnote{\href{https://github.com/AndresdPM/BANNJOS}{https://github.com/AndresdPM/BANNJOS}, \href{https://gitlab.cefca.es/adelpino/bannjos}{https://gitlab.cefca.es/adelpino/bannjos}} machine learning pipeline designed to derive any desired property of an astronomical object using supervised learning techniques. It is coded in \texttt{Python} using the \texttt{Tensorflow} and \texttt{Tensorflow\ probability} libraries \citep{tensorflow15}. \bannjos{} works as a general-purpose regressor and is designed to operate at a variety of user input levels. It can run nearly automatically once provided with the minimum input files, but allows for extensive customization through optional keywords, giving users greater control over the process.

In essence, \bannjos{} trains a model, \(f(\mathbf{x})\), based on the relation between a dependent variable \(y_{\mathrm{true}}\) and a set of independent variables, \(\mathbf{x}\), using a training sample. It then predicts this variable, \(y^\prime_{\mathrm{pred}}\), in any given \(\mathbf{x}^\prime\). Its most critical components include:

\begin{itemize}
    \item Reading and preprocessing the data,
    \item Training the model,
    \item Computing \(y_{\mathrm{pred}}\).
\end{itemize}

A flowchart showcasing the working flow of \bannjos{} can be found in Figure~\ref{fig:Bannjos_workflow}. In the following we explain the details of the preprocessing of the data and the different models available.

\subsection{Preprocessing of the Data}\label{sec:BANNJOS_preprocessing}

The training sample consists of a table containing $\mathbf{x}$ and $y_\mathrm{true}$. The nature of these variable sets may vary; for instance, $y_\mathrm{true}$ can include categorical, continuous, or both types of information, whereas $\mathbf{x}$ may encompass diverse types of data, such as photometric magnitudes or the positions of sources on the detector's focal plane. Adapting all available information for the training process is a critical step with a notable impact on the final results. Initially, \bannjos{} shuffles the training set and constructs $\mathbf{x}$ and $y_\mathrm{true}$ based on user-defined options and criteria. The data are then divided into training and test sets, with the latter containing a subset of the training data that will not be exposed to the model during its training. This test sample, denoted as $\mathbf{x}^\prime$ and $y_\mathrm{true}^\prime$ in Figure~\ref{fig:Bannjos_workflow}, will serve as a validation sample to assess the quality of the results.

Next, \bannjos{} normalizes $\mathbf{x}$, an essential step prior to training a neural network. Outliers can artificially widen the dynamic range of the training sample, thereby compressing the range containing actual information. To mitigate this, normalization is performed individually to each variable using the 0.005 and 0.995 quantiles of $\mathbf{x}$ as the minimum and maximum, respectively, and then clipping the normalized set to the range [0, 1]. Any missing data in $\mathbf{x}$ are assigned a default value of -0.1. This approach allows \bannjos{} to handle missing data within the training sample. Lastly, \bannjos{} converts any categorical data into continuous numerical values, enabling it to treat classification problems as regression problems by assigning probabilities to each class.

\subsection{Dealing with Uncertainties with Different Models}\label{sec:BANNJOS_models}

\bannjos{} provides users the option to choose between six different regression models: three deterministic and three probabilistic. The deterministic models include a k-neighbors (KNR), a random forest (RFR), and a multilayer feed-forward artificial neural network (ANN). The probabilistic models consist of an ANN with a Gaussian posterior and two Bayesian ANNs (BANNs): a variational inference ANN and a dropout ANN.

Each model has its own set of advantages and disadvantages. Simple models like $k$-neighbors and random forest regressors are highly scalable and perform quickly on large, well-distributed samples, with few hyperparameters requiring tuning. However, they may not be the best choice for highly complex problems and are incapable of extrapolating data. ANNs based regressors may provide better results for complex problems but require more hyperparameters to be fine-tuned and demand greater computational resources.

Measuring reliable uncertainties associated with each model's predictions is crucial for most scientific cases. \bannjos{} computes these uncertainties, $\sigma(y_\mathrm{pred})$, in different ways depending on the model's capabilities.

\subsubsection{Deterministic Models}

Three from the six models implemented in \bannjos{} are unable of predicting the uncertainty associated with a prediction's nominal value. These models include KNR, RFR, and the basic ANN. In such cases, \bannjos{} calculates the expected uncertainty using a $k$-fold cross-validation scheme. In brief, the training sample is randomly divided into $k$ equal-sized subsamples. A nominal model is then trained on $k-1$ subsamples and used to predict values in the remaining subsample, which acts as the validation data. This process is repeated $k$ times\footnote{The $k$ value is set by the user. Larger $k$ values will reduce bias in the prediction, but at larger computational cost.}, with each of the $k$ subsamples used once as validation data. After completing the $k$-fold cross-validation, the results are used by \bannjos{} to train a second model (the variance model) based on the relation between $\left|y_\mathrm{true} - y_\mathrm{pred}\right|^2$ and $\mathbf{x}$. This variance model predicts the uncertainties for the nominal model's predictions. In our trials, this method proved to be faster than using $k$ models trained during the $k$-fold cross-validation to predict uncertainty, specially when predicting in large data sets, while providing very similar results.

\subsubsection{Probabilistic Models}

\bannjos{} includes three probabilistic neural network models, two of which operate as Bayesian approximations. The first model, an ANN with a Gaussian posterior, fits and predicts both the nominal result and its aleatoric uncertainty, i.e., the uncertainty associated with randomness observed in the training sample. The variational inference BANN and dropout BANN models, however, can estimate both aleatoric and epistemic uncertainties, the latter arising from insufficient information in the training sample.

If a probabilistic model is used, $y_\mathrm{pred}$ and its associated uncertainty are derived by sampling the model's posterior multiple times ($N$). This process is equivalent to sampling the PDF of the prediction for each instance (astronomical object), $y_\mathrm{pred, i}$. \bannjos{} computes several useful statistics from these PDFs sucha as the mean and several percentiles (see Section~\ref{sec:training_sampling} and Appendices~\ref{Appendix:Data_compression}, and ~\ref{Appendix:Output}).

\section{Using \bannjos{} on J-PLUS for Object Classification}\label{sec:bannjos_on_JPLUS}

In this work, we have utilized \bannjos{} for object classification in J-PLUS. We adopted a systematic approach to construct the training sample, select the optimal model and its corresponding hyperparameters, and derive the probability distribution functions, PDFs, for each object's likelihood of belonging to a specific class.

\subsection{Training Sample}\label{sec:Training_section}

\begin{figure*}
\begin{center}
\includegraphics[width=\linewidth]{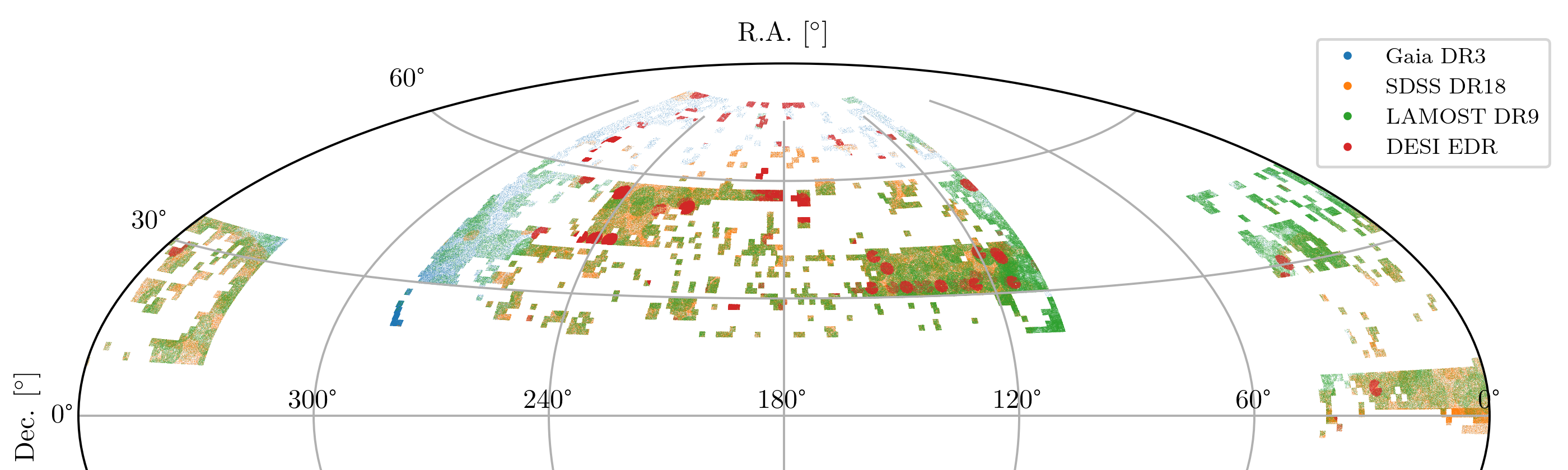}
\caption{Aitoff projection of the training list sources. The positions of the sources are represented by small dots, color-coded according to their originating survey. These are sources identified in J-PLUS with available spectroscopic or photospectroscopic classification. The small squares with 1.4-degree sides reflect the J-PLUS tiles. For a reference on the number of sources, see Table~\ref{tab:contributions}.}
\label{fig:Aitoff_training}
\end{center}
\end{figure*}

We compiled an extensive training sample consisting of a selection of objects with J-PLUS photometry and available spectroscopic classification. We first downloaded the entire J-PLUS catalog, with over 47 million sources, taking information from several of the main scientific tables available at CEFCA's archive\footnote{\href{https://archive.cefca.es/catalogues/jplus-dr3}{https://archive.cefca.es/catalogues/jplus-dr3}}. We combined the resulting table with data from ongoing and past all-sky surveys that contain additional information about the observed sources. These data include astrometric and photometric information from \Gaia{} DR3 \citep{gaia_mission, gaia_dr3} and \Catwise{} \citep{catwise2020}, as well as reddening information from dust maps of \citet{schlafly11}. In order combine all the data, we took advantage of the catalogues matches already available at the J-PLUS archive. Details on the specific J-PLUS archive query and the variables utilized during the training are available in Appendix~\ref{Appendix:Data_query}. Broadly, the training list encompasses data on:

\begin{itemize}
    \item J-PLUS photometry, i.e., fluxes measured across eight apertures for the 12 bands using forced photometry (\texttt{SExtractor} dual mode);
    \item J-PLUS position on the CCD and morphology, i.e., ellipticity, effective radius;
    \item J-PLUS photometry and masking flags;
    \item J-PLUS Tile observation details, i.e., seeing, zero-point, noise, etc;
    \item \Catwise{} photometry, i.e., \texttt{W1mpro\_pm} and \texttt{W2mpro\_pm} bands;
    \item \Gaia{} and \Catwise{} astrometry, i.e., parallaxes and the absolute value of the one-dimensional proper motions\footnote{Two-dimensional proper motions could potentially leak positional information (RA, Dec) to the model due to the Solar Galactic Reflex Velocity.}.
\end{itemize}

Flux measurements across different apertures provide additional insights into the source's morphology. Its position in the detector can also be relevant, as it may help identify geometric distortions or other image-affecting cosmetic effects. Furthermore, details regarding the tile image's quality, such as average seeing during acquisition, can be crucial for determining whether the source is extended. In total, we included 445 variables in our analysis.

In order to obtain the spectroscopic classification, we performed a cross-match of the resultant J-PLUS+\Gaia{}+\Catwise{} table with the {\tt SpecObj} table from \SDSS{} DR18 \citep{sdss_dr18}, the Low Resolution spectroscopy catalog from \LAMOST{} DR9 (Large Sky Area Multi-Object Fiber Spectroscopic Telescope; \citealt{lamost1}), and the Early Data Release (EDR) from \DESI{} (Dark Energy Spectroscopic Instrument; \citealt{desi_edr}). The match is based on the sky position of sources across the catalogs, considering a source identical if its registered positions differ between catalogs by less than a specified angular separation. This maximum separation was determined in two steps. In the first step, a very large radius of 2 arcsec was used, ensuring all possible matches were included. In the second step, the maximum separation was determined based on finding 99\% of matches within it. For J-PLUS and \SDSS{}, the maximum allowed separation was found to be 0.6 arcseconds, while for J-PLUS with \LAMOST{} and \DESI{}, it was set to 0.65 arcseconds. Lastly, we included a set of unequivocally classified \Gaia{} sources, where the probability of belonging to a particular class is 1 and 0 for the other two classes. The resulting catalog underwent cleaning and consistency checks. Initially, we filtered out poorly measured sources by applying recommended criteria from SDSS:

\begin{itemize}
    \item $0.864 < ${\tt RCHI2} $< 1.496$,
    \item {\tt Z\_ERR} $> 0$,
    \item {\tt Z\_ERR}/(1+{\tt Z}) $ < 3\times10^{-4}$,
    \item {\tt PLATEQUALITY} not {\tt "bad"},
    \item ({\tt SPECPRIMARY} = 1) or ({\tt SPECLEGACY} = 1),
    \item {\tt SN\_MEDIAN\_ALL} $ \geq 2$,
    \item {\tt ZWARNING\_NOQSO} $ = 0$.
\end{itemize}

The clipping levels for {\tt RCHI2} were set to the 0.1 and 0.9 quantiles of its distribution, respectively. For sources from \LAMOST{}, the following criteria were applied:

\begin{itemize}
    \item {\tt FIBERMASK} $ = 0$,
    \item {\tt Z\_ERR} $> 0$,
    \item {\tt Z\_ERR}/(1+{\tt Z}) $ < 3\times10^{-4}$,
    \item {\tt SN\_MEDIAN\_ALL} $ \geq 3$.
\end{itemize}

For the reliable sources classified in \DESI{} EDR, we selected those with:

\begin{itemize}
    \item {\tt TARGETID} $> 0$,
    \item {\tt SV\_PRIMARY} = 1,
    \item {\tt ZWARN} $ = 0$.
\end{itemize}

\begin{figure}
\begin{center}
\includegraphics[width=\linewidth]{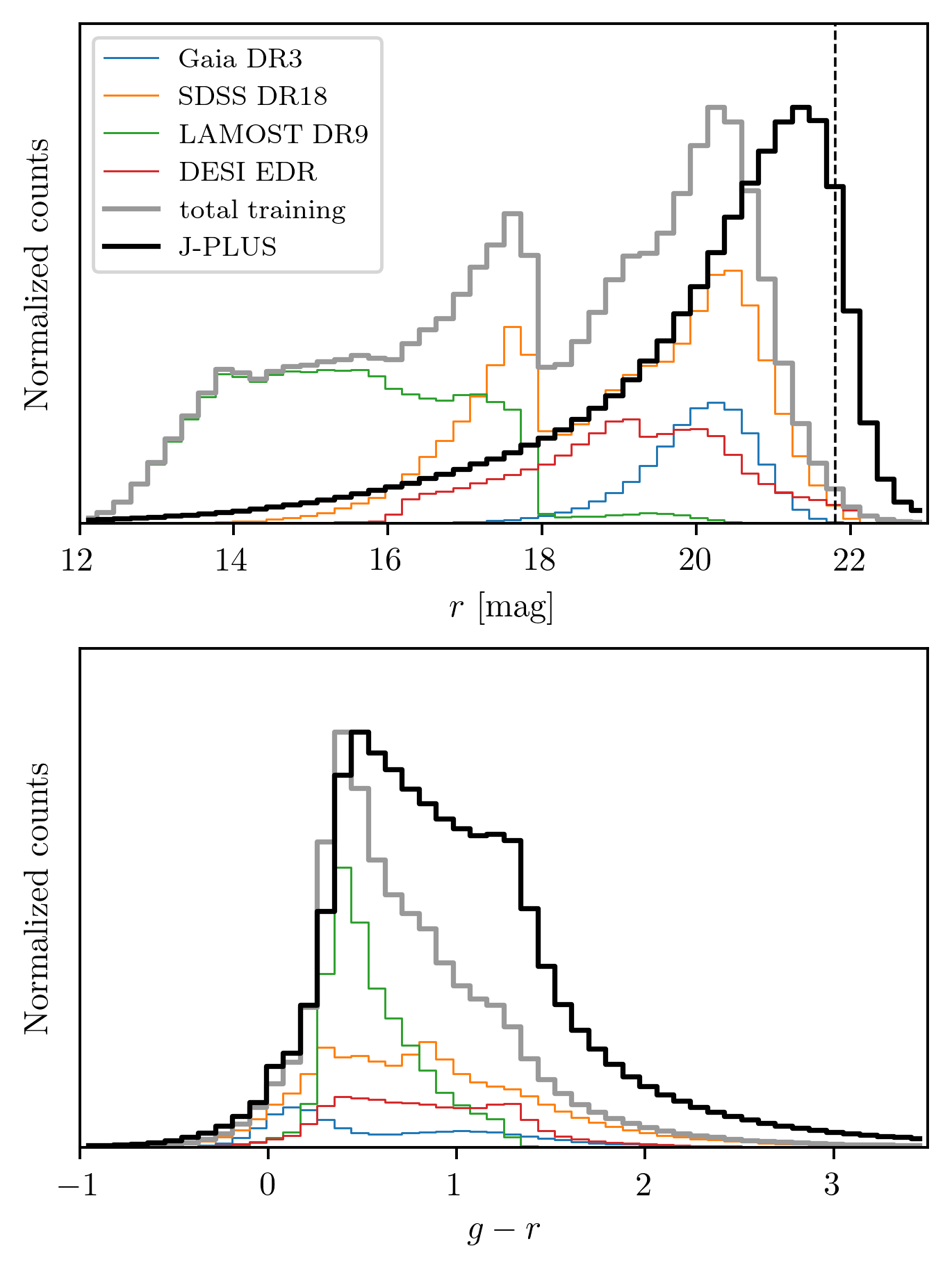}
\caption{Distribution of the $r$ magnitude (top panel) and $g-r$ color (bottom panel) for the sources in J-PLUS and the training list. The distribution of sources from each contributing survey is depicted by thin lines in various colors (as in Figure~\ref{fig:Aitoff_training}). The aggregate training list is represented by a thick gray line, while the entirety of J-PLUS sources, 47,751,865 objects, is shown by a thick black line. Histograms related to the training list have been normalized to the peak of the total training list (gray line), whereas the J-PLUS histograms have been normalized to their own peak. A vertical thin dashed line shows the limiting magnitude of J-PLUS DR3 ($r = 21.8$ mag, see Table~\ref{tab:JPLUS_filters}) The extended tail of objects with $g-r \gtrsim 1.5$ is mostly composed by low S/N sources with poorly determined $g$ magnitude.}
\label{fig:Mag_color_training}
\end{center}
\end{figure}

\begin{figure*}
\begin{center}
\includegraphics[width=\linewidth]{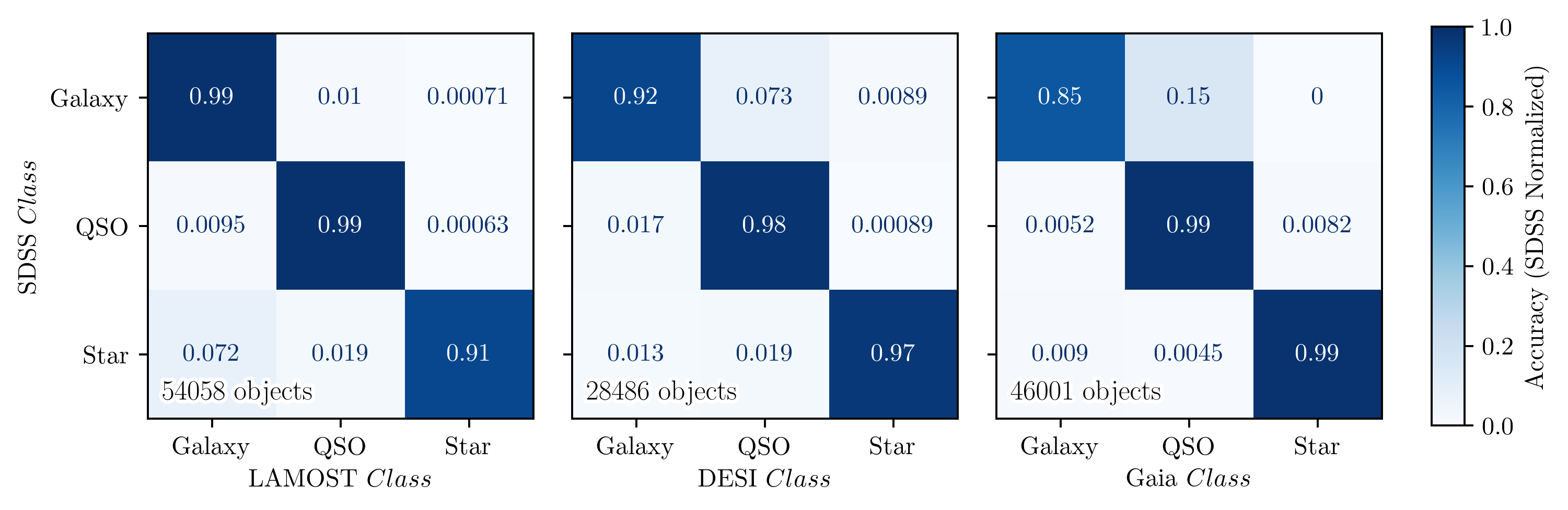}
\caption{Confusion matrices between the four surveys used to compile the training list. The proportion of objects in each bin relative to the total number of objects in \SDSS{} is indicated by varying shades of blue and is presented as a real number $\in [0, 1]$. The confusion matrices also specify the total number of objects used in their computation.}
\label{fig:Confusion_training}
\end{center}
\end{figure*}

Lastly, \Gaia{} sources were chosen based solely on their renormalized unit weight error, with {\tt RUWE} $\leq 1.3$. The applied quality filters resulted in 586,641 sources from \SDSS{}, 508,432 from \LAMOST{}, 256,532 from \DESI{}, and 152,566 from \Gaia{}, totaling 1,504,171 sources. 

The spatial distribution of these sources is illustrated in Figure~\ref{fig:Aitoff_training}. Histograms showing the magnitude and color ranges covered by each of the surveys in the training list are displayed in Figure~\ref{fig:Mag_color_training}. Although the training list spans the entire J-PLUS footprint, the levels of completeness and color coverage vary, inherent to each contributing survey. For instance, \Gaia{} (depicted in blue) provides uniform sampling across the footprint but is limited to relatively blue sources ($g-r \lesssim 1.6$) and does not reach the depths of its spectroscopic counterparts. Conversely, surveys such as \SDSS{} or \DESI{} delve deeper and sample the less populated galactic poles, offering insights into distant galaxies and QSOs. Incorporating various catalogs into the training list maximizes the coverage across the parameter space spanned by J-PLUS, minimizing instances where the model must extrapolate predictions, which, in theory, should enhance results. However, due to the disparate coverage in terms of object types, photometric properties, and spatial distribution among the surveys, it is crucial to exclude any positional information from the training list to prevent the model from learning the inherent biases in the selection and properties of the sources inherent to spectroscopic surveys.

A significant number of sources have multiple measurements in two or more of the four classification catalogs. Confusion matrices for the classification of such objects, shown in Figure~\ref{fig:Confusion_training}, illustrate the consistency (or lack thereof) between the surveys, indicating the relative number of objects classified identically or differently between pairs of surveys. These matrices were constructed using all objects shared between each survey, with \SDSS{} serving as the reference. Although inter-survey consistency is generally good, it is not without discrepancies. For example, \LAMOST{} tends to classify fewer objects as stars compared to \SDSS{} (91\%), and \Gaia{} often classifies many of \SDSS{}'s galaxies as QSOs (14\%). Despite these differences, the overall agreement is satisfactory, especially considering that \Gaia{} contributes only around 9\% of the galaxies and QSOs to the final list.

We removed all but one instance from repeated objects with consistent spectroscopic classes across catalogs (1160 in total), retaining duplicates where classifications differed depending on the catalog. We identified a small subset of objects (747) with varying classifications across two or more input lists. After visually inspecting the spectra of some of these objects, we found they were primarily galaxies misidentified as stars and vice versa (233 cases), and distant active galaxies classified either as QSOs or galaxies (467 cases). We refined this last group further by excluding nearby sources (stars) with parallax $\parallax/\sigma(\parallax) > 1$, meaning objects with parallax different from zero at a 1$\sigma$ confidence level. Since the dichotomy between galaxies and QSOs is not always clear, we decided to keep those objects (426), but removed those with implausible class combinations, such as Galaxy-Star or QSO-Star. Notably, we found no objects classified differently across three or more catalogs, meaning no objects were assigned all three possible classes.

The final list comprises 1,365,700 objects (1,365,274 unique), distributed into 480,267 galaxies, 127,633 QSOs, and 757,800 stars. Each object's record in the catalog includes photometric, astrometric, and morphological data from J-PLUS, \Gaia{} DR3, and \Catwise{}, alongside spectroscopic classification from \SDSS{} DR18 (585,336 objects in common), \LAMOST{} DR8 (507,737 objects in common), \DESI{} EDR (254,586 objects in common), and spectrophotometric classifications from \Gaia{} DR3 (151,974 objects in common). The details about the composition of the training set can be found in Table~\ref{tab:contributions}. We reserved 10\% of these sources to compose the 'test' set, which will be utilized later to validate and assess our results.

\begin{table}
\begin{small}
    \centering
    \caption{Composition of the training set.}
    \label{tab:contributions}
    \begin{tabular}{lccc|c}
        \hline\hline\rule{0pt}{3ex}
        Survey & Galaxies & QSOs & Stars & Total\\
        \hline 
        \SDSS{} DR18 & 354,849 & 93,905 & 137,887 & 586,641 \\
        \LAMOST{} DR8 & 28,289 & 11,219 & 468,924 & 508,432 \\
        \DESI{} EDR & 140,604 & 15,029 & 100,899 & 256,532  \\
        \Gaia{} DR3 & 1,014 & 74,391 & 77,161 & 152,566 \\
        \hline 
        Total & 524,756  & 194544 & 784,871  & 1,504,171  \\
        Total non-repeat\tablefootmark{a} & 480,267 & 127,633 & 757,800 & 1,365,700 \\
        \hline\hline
        Training no balan. & 432,315 & 114,774 & 682,041 & 1,229,130 \\
        Balan. downsamp.\tablefootmark{b} & 114,774 & 114,774 & 114,774 & 344,322 \\
        Balan. supersamp.\tablefootmark{c} & 682,041 & 682,041 & 682,041 & 2,046,123 \\
    \end{tabular}
\end{small}
\tablefoot{
\tablefoottext{a} {Total non-repeated sources. Counts after removing repeated sources present in one or more surveys.}
\tablefoottext{b} {Balanced training list down-sampled to the least common class.}
\tablefoottext{c} {Balanced training list super-sampled to the most common class.}
}
\end{table}

The compiled training list is disproportionately biased towards the "star" class, with galaxies being the second most prevalent. Indeed, within the J-PLUS footprint, stars are the most numerous due to the survey's depth limitations, but the class ratios in the training list might be artificially skewed by the design of the contributing surveys. To evaluate and mitigate potential biases, we tested our models using three different versions of the training list. The first version retained the original composition after removing 10\% for the test sample, resulting in 1,229,130 sources. The second, a downsampled version, randomly reduced the numbers of the two most abundant classes to match those of the least populated class, the QSOs, resulting in a list of 344,322 equally distributed sources among the "Star", "QSO", and "Galaxy" classes. The third, a balanced version of the original list, employed oversampling for the less populated classes using a Synthetic Minority Over-sampling Technique (SMOTE) (\citealt{smote}), which uses a k-neighbor interpolator to generate new instances (sources) based on the averages of neighboring properties. This "augmented" list, balanced among classes, contains 2,046,123 sources.

\subsection{Model Selection and Hyperparameter Tuning}\label{sec:Model_Selection}

To achieve optimal classification, we evaluated the models described in Section~\ref{sec:BANNJOS_models}, identifying the model and hyperparameter configuration yielding the best results. Additionally, we assessed the three variations of our training sample outlined in Section~\ref{sec:Training_section}. The performance of each configuration was measured on the test sample through cross-validation, comparing the spectroscopic/spectrophotometric class of an object, \(y_{\mathrm{true}}\), against the predicted one, \(y_{\mathrm{pred}}\). For BANNs, which provide the PDF for each class, we designated the class corresponding to the highest median value (quantile 0.5) from the three PDFs as the predicted class  \(class = \max \left[ PC_{\text{Galaxy}}(50), PC_{\text{QSO}}(50), PC_{\text{Star}}(50)\right]\). We started evaluating the models in a coarse grid of hyperparameters in order to gain a general idea about their performance. All the ANN based models were trained using a validation sample of 40\%\footnote{The model sets apart this fraction of the training list, not training on it, and evaluates the loss metrics on this data at the end of each epoch (training iteration).} In general, ANNs significantly surpassed RFR and KNR in performance, with a deep dropout BANN emerging as the superior model. This result is somewhat expected, as dropout BANNs are less susceptible to overfitting compared to traditional and variational inference ANNs, especially when the model architecture is sufficiently deep. With the exception of RFR, the augmented training list consistently yielded slightly better results across all models. After selecting the dropout BANN as the best overall model, we evaluated its performance across an extensive grid of potential hyperparameter configurations:

\begin{itemize}
    \item {\tt Number of hidden layers}: $[8, 7, 6, 5, 4, 3, 2]$
    \item {\tt Batch size}: $[32, 64, 128, 256, 512]$
    \item $L_n$: $[1600, 1300, 1000, 700, 500, 300, 200, 100, 50, 5]$
    \item {\tt Dropout ratio at $L_{1-8}$}: $[0.1, 0.2, 0.3, 0.4, 0.5]$
    \item {\tt Dropout ratio at $L_0$}: $[0.0, 0.1, 0.2]$
    \item {\tt Loss function}: MSE, Huber
    \item {\tt Initial learning rate}: $[10^{-3}, 5 \times 10^{-4}, 10^{-4}, 5 \times 10^{-5}]$
    \item {\tt Step decay in learning rate}: $[10, 20, 30, 50, \infty]$
\end{itemize}

Here, {\tt Number of hidden layers} denotes the quantity of hidden layers between the input and output. {\tt Batch size} refers to the number of samples processed before updating the model. The term $L_n$ indicates the number of neurons in the $n$-th layer for $1 \leq n \leq 8$. The {\tt Dropout ratios} represent the fraction of randomly dropped neurons in each epoch at hidden layers ($n > 0$) or at the input layer ($n = 0$). The {\tt Loss function} is the metric minimized during the fitting process. {\tt Initial learning rate} denotes the step size at each epoch during optimization, and {\tt Step decay in learning rate} specifies the epoch count before the learning rate updates to half of its previous value, with $\infty$ indicating a constant learning rate.

Given the vast number of possible hyperparameter combinations, exceeding $10^{11}$, testing all configurations was computationally unfeasible. Therefore, we employed a Random Search Cross-Validation approach, randomly selecting and testing 800 configurations against the test sample. We limited each model's training to 2000 epochs and incorporated an early stopping mechanism that halts training if the loss does not improve over 50 epochs\footnote{Improvement is defined within a precision of $10^{-5}$ in the loss function.}. To make the predictions, we sampled the posterior only 128 times. This allows us to keep a reasonable computational cost for all the optimization tests, while realistically sampling the hyperparameter space. Upon completing cross-validation for the 800 models, we utilized a Histogram-based Gradient Boosting Regression Tree to identify the optimal model architecture and training hyperparameters. Model evaluation was based on three metrics: balanced accuracy, average precision, and their quadratic sum. The found optimal hyperparameters for the dropout BANN are:

\begin{itemize}
    \item {\tt Number of hidden layers}: 4
    \item {\tt Batch size}: 256
    \item \(L_{1-4} = [700, 1300, 500, 300]\)
    \item {\tt Dropout ratio at \(L_{1-4}\)}: 0.4
    \item {\tt Dropout ratio at \(L_0\)}: 0
    \item {\tt Loss function}: MSE
    \item {\tt Initial learning rate}: \(10^{-4}\)
    \item {\tt Step decay in learning rate}: 30
\end{itemize}

Where now $L_{1-4}$ is the number of neurons in the hidden layers 1 to 4. Despite this being the best configuration, our testing indicates that model performance is largely invariant to hyperparameter configuration, provided the values are reasonable. For instance, the accuracies of our top 10 models differ by less than 0.1\%. More details on the impact on the accuracy that the model hyperparameters have, the correlation between some hyperparamenters, and the Individual Conditional Expectation (ICE) curves, can be found in Appendix~\ref{Appendix:Model_Selection}.

\subsection{Training the Model and Sampling the Posterior}\label{sec:training_sampling}

\begin{figure*}
\begin{center}
\includegraphics[width=\linewidth]{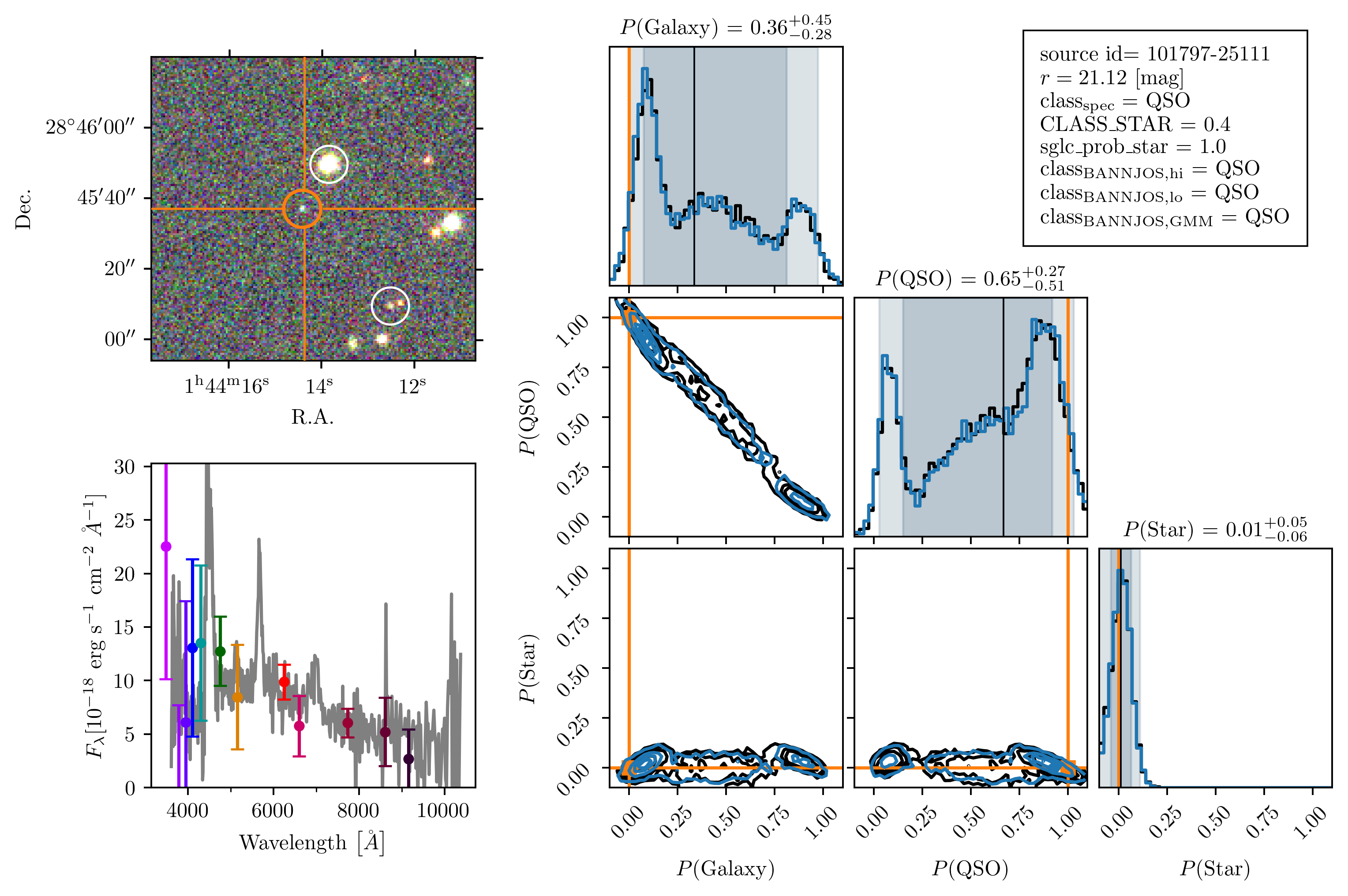}
\caption{Example of results obtained with \bannjos{} for the source {\tt Tile Id} = 101797, {\tt Number} = 25111. {\it Upper left:} J-PLUS color image composed using the $r$, $g$, and $i$ bands, normalized between the 1st and 99th percentile of the total flux. The classified source is at the center of the reticle, marked by an orange open circle, while other sources detected in the J-PLUS catalog are marked with white open circles. {\it Lower left:} SDSS spectrum undersampled by a factor of 10 for improved visibility (in gray), alongside J-PLUS photometry across its 12 bands. {\it Right:} Corner plot illustrating the three-dimensional posterior probability distributions for $P(\text{class} = \text{Galaxy})$, $P(\text{class} = \text{QSO})$, and $P(\text{class} = \text{Star})$. Orange lines denote the spectroscopic class, $y_\mathrm{true}$. Black contours and histograms represent the posterior probability distribution sampled 5000 times ($\text{class}_{\text{BANNJOS,hi}}$), with black lines indicating the median probability for each class and the gray-shaded area covering the 2nd to 98th (lighter gray) and the 16th to 84th (darker gray) percentile ranges. Blue contours, histograms, and shaded areas depict the reconstructed posterior probability distribution from the GMM model fitted to $N=300$ points. A text box in the upper right corner lists the complete source ID, its $r$ magnitude, the 'true' classification from spectroscopy ($\text{class}_{\text{spec}}$), its \texttt{CLASS\_STAR} and \texttt{sglc\_prob\_star} scores, and the \bannjos{} classifications from the high-quality posterior sampling ($N=5000$, $\text{class}_{\text{BANNJOS, hi}}$), the regular-quality posterior sampling ($N=300$, $\text{class}_{\text{BANNJOS, lo}}$), and the classification from the reconstructed PDF following the GMM compression method, $\text{class}_{\text{BANNJOS,GMM}}$. The classification is determined as the one with the highest median probability value. Despite its complexity, the PDFs obtained from sampling 5000 times and the one after reconstruction from the GMM are nearly indistinguishable, with gray and blue contours and shaded areas covering the same areas. The corresponding classifications, $\text{class}_{\text{BANNJOS,hi}}$ (black) and $\text{class}_{\text{BANNJOS,GMM}}$ (blue), also match the spectroscopic classification. The GMM compression procedure is described in Appendix~\ref{Appendix:Data_compression}.}
\label{fig:Individual_1}
\end{center}
\end{figure*}

\begin{figure*}
\begin{center}
\includegraphics[width=\linewidth]{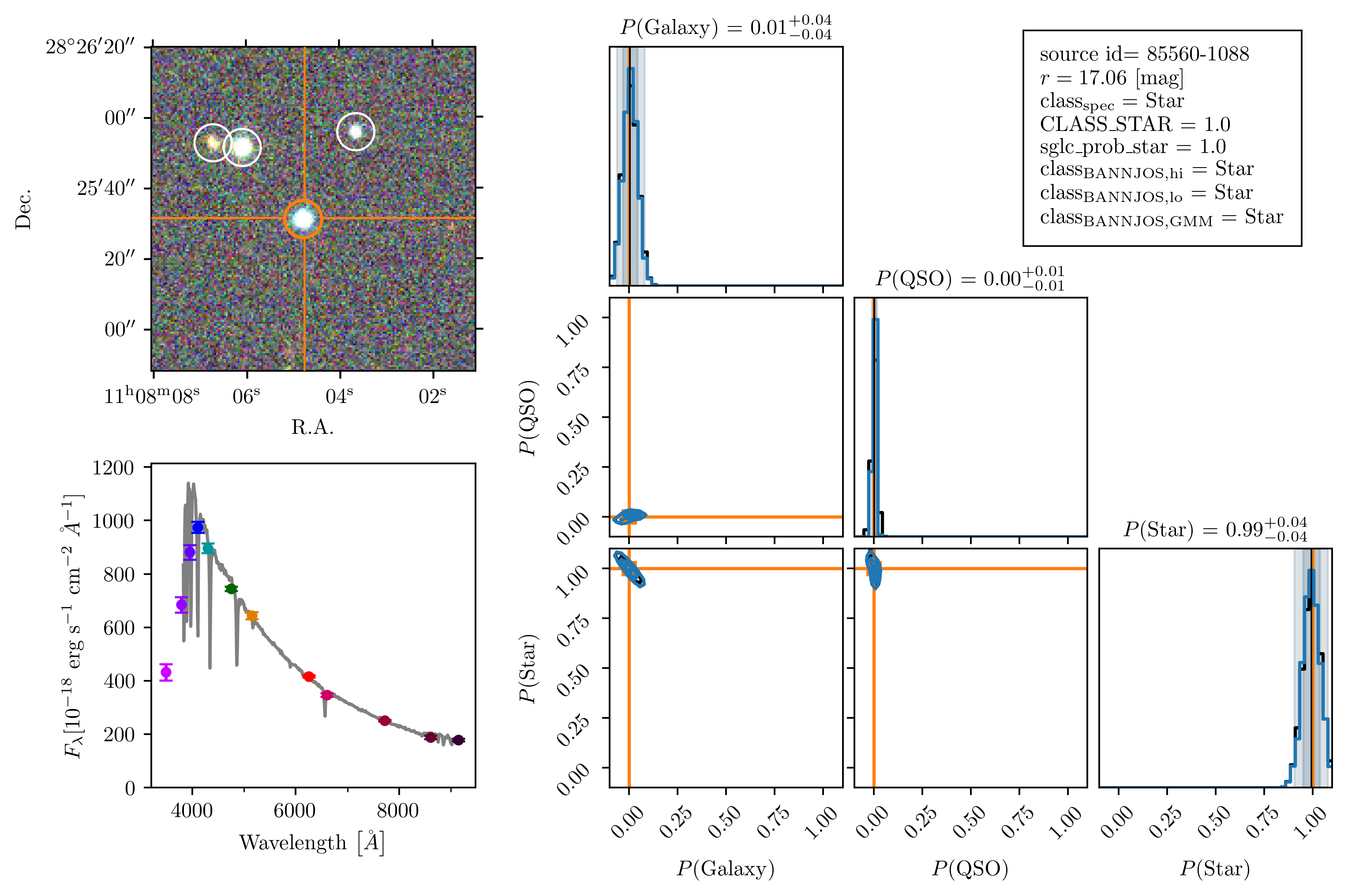}
\caption{Example of results obtained with \bannjos{} for the source {\tt Tile Id} = 85560, {\tt Number} = 1088. At $r= 17.06$ mag, \bannjos{} classifies this source as a star with high confidence. The markers correspond with those used in Figure~\ref{fig:Individual_1}. Most sources classified with \bannjos{} will show PDFs similar to this one, with very little dispersion around the predicted class.}
\label{fig:Individual_2}
\end{center}
\end{figure*}

We used the hyperparameters derived in the previous Section~\ref{sec:Model_Selection} to build our model. To maximize results, we extended the training stopping criteria to 10,000 epochs and updated the early stop function to 500 epochs. The model converged after 8,963 epochs with a loss function precision of \(10^{-7}\).

A significant advantage of using a BANN model is its capability to provide the full PDF for \(y_{\mathrm{pred}}\). Here, the total PDF for a specific object is the sum of the individual class PDFs, defined within the 3-dimensional space delineated by orthogonal axes \(P(\text{class} = \text{Galaxy})\), \(P(\text{class} = \text{QSO})\), and \(P(\text{class} = \text{Star})\). Sampling the model multiple times, \(N\), in a Monte-Carlo fashion, \bannjos{} outputs three probabilities for each sample, corresponding to the likelihood of the object belonging to each class. Due to the model's stochastic nature, different samples yield unique outcomes, resulting in \(N \times 3\) data points. Figure~\ref{fig:3D_Distro} shows a sampling example with $N = 5000$. 

While maintaining and analyzing all points permits comprehensive PDF reconstruction and facilitates informed class determination, extensive posterior sampling (large \(N\)) is computationally intensive and data-heavy, becoming impractical for the entire J-PLUS catalog. To alleviate computation time and storage, we employed a reduced sampling count (\(N = 300\), denoted as \(\text{class}_{\text{BANNJOS, lo}}\)) and projected the 3-dimensional probabilities onto the plane defined by \(P(\text{class} = \text{Galaxy}) + P(\text{class} = \text{QSO}) + P(\text{class} = \text{Star}) = 1\), thereby reducing the dimensionality to two. We then applied a 2-dimensional Gaussian Mixture Model (GMM) with three components to model the projected probabilities. The GMM parameters—covariance matrices, means, and weights—are compactly stored, effectively capturing the posterior's essence with significantly fewer parameters. \bannjos{} additionally computes mean, median absolute deviation (MAD), and specified percentiles for each class's cumulative distribution, along with Pearson's correlation coefficients between classes. The compression method and its efficacy are explained in more detail in Appendix~\ref{Appendix:Data_compression}.

In Figure~\ref{fig:Individual_1}, we showcase a faint source (\(r = 21.12\) mag) classified by \bannjos. Due to its high photometric and astrometric uncertainties, \bannjos{} assigns a low-confidence QSO classification to the source with \(P(\text{class} = \text{QSO}) = 0.65^{+0.27}_{-0.51}\). The derived PDF exhibits multiple peaks and a pronounced anticorrelation between \(P(\text{class} = \text{QSO})\) and \(P(\text{class} = \text{Galaxy})\), indicating a considerable probability that the object could also be a galaxy with $P(class = {\rm Galaxy}) = 0.36^{+0.45}_{-0.28}$. This ambiguity underscores the intricate nature of distinguishing between QSOs and active galaxies, particularly when faced with unresolved targets. Indeed, when checking the \SDSS's \code{SUBCLASS} field for this an other similar objects, we found all to be unresolved active galaxies or QSOs. Interestingly, in this particular case the \code{sglc\_prob\_star} classifier wrongly assigned a very high probability of being a star to the object, showcasing how the combination of photometric, astrometric and morphological information used by \bannjos{} can help to distinguish sources based on their nature, rather than just their apparent aspect, even in low S/N conditions.

The implemented GMM compression method was tested against an additional validation sample with $N = 5000$ BANN posterior samples ($\rm class_{BANNJOS, hi}$). The GMM faithfully reproduces the original results. The high-quality sampling (black histograms in Figure~\ref{fig:Individual_1}), and the ones using $N=300$ compressed with the GMM ($\rm class_{BANNJOS, GMM}$, blue histograms in Figure~\ref{fig:Individual_1}) are consistent across the entire test sample, barring rare exceptions. Only in 4 objects with very low S/N (0.003\% of the cases, see Appendix~\ref{Appendix:Data_compression}), we obtained a different classification using the highest median probability of the three classes, indicating that the implemented compression method provides a reliable classification. 

In contrast with the low S/N example, in Figure~\ref{fig:Individual_2} we show an example of a source classified by \bannjos{} with high-confidence. The relatively bright source ($r = 17.06$ mag), is unequivocally identified as a star by \bannjos{}: $P(class = {\rm Star}) = 0.99^{+0.04}_{-0.04}$, $P(class = {\rm Galaxy}) = 0.01^{+0.04}_{-0.04}$, $P(class = {\rm QSO}) = 0.00^{+0.01}_{-0.01}$. Most sources brighter than $r \sim 20$ mag will show posteriors that tightly distribute around the predicted class with very little dispersion or correlation to other classes probabilities.

\section{Validation of the Model}\label{sec:Validation_model}

\begin{figure*}
\begin{center}
\includegraphics[width=\linewidth]{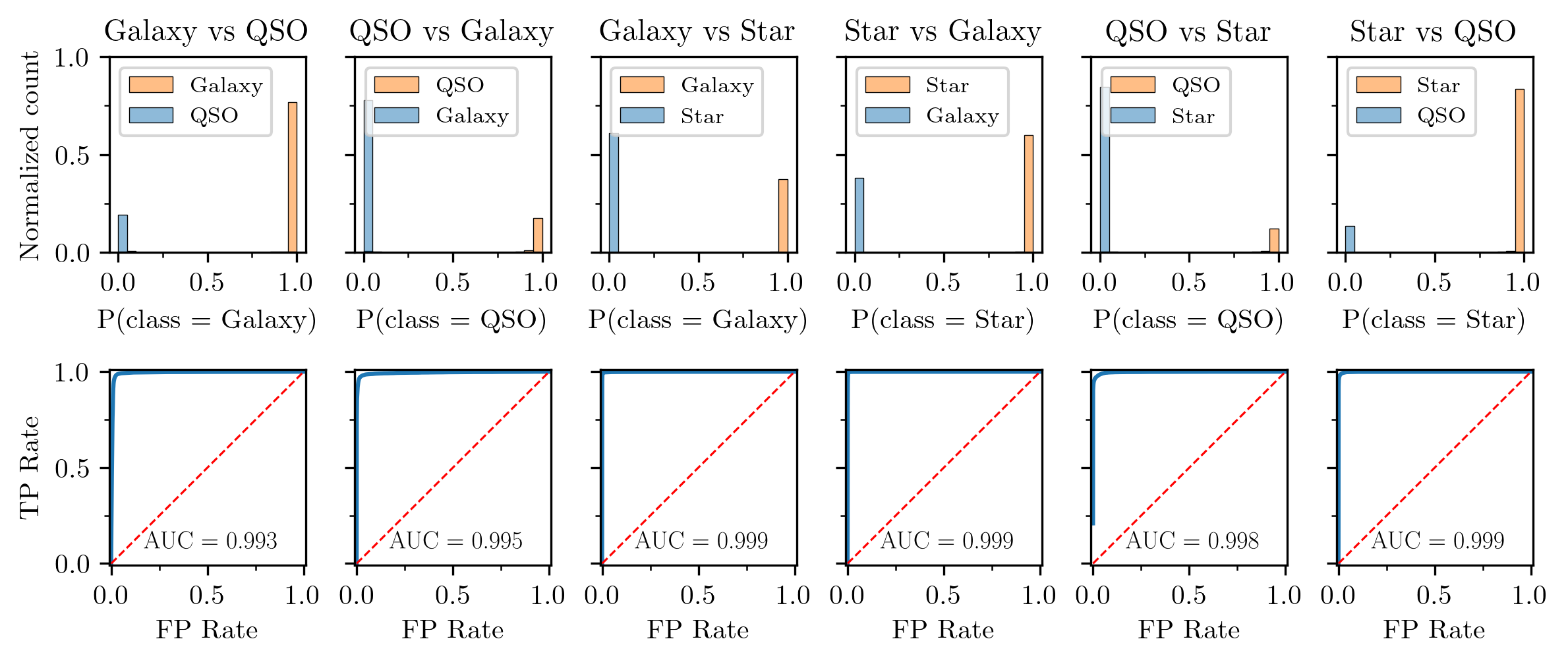}
\caption{Model performance evaluated up to $r = 21.9$ mag, the median completeness for compact sources in J-PLUS, assessed on the test sample consisting of 136,570 objects. The top panels illustrate the distribution of probabilities for objects being classified according to their spectroscopic category. The difference in bar heights reflex the different amount of sources present in the test sample. The bottom panels depict the Receiver Operating Characteristic (ROC) curves in blue for each class combination. These curves approach the maximum True Positive Rate (1) almost immediately, demonstrating excellent model performance. The Area Under the Curve (AUC) exceeds 0.99 for all six class combinations.}
\label{fig:ROC_curves}
\end{center}
\end{figure*}

In this section, we validate the classification outcomes of \bannjos{} using the identical test sample and criteria established in Section~\ref{sec:Model_Selection}. This involves assigning the predicted class of the object, \(y_{\mathrm{pred}}\), and comparing these with the spectroscopic classifications, \(y_{\mathrm{true}}\). Since \bannjos{} provides the full PDF of the classification, there are multiple criteria that could be used in order to assign a class to each source. However, we will focus now on the most simple approach that is assigning classes based on the highest median probability value across the three PDFs. In Section~\ref{sec:Refining_selection}, we explore how the purity of the selected samples can be improved using more sophisticated selection criteria. It is worth mentioning that the test sample was subtracted from the training list before any preprocesing of the data, and thus it is unbalanced with proportions very similar to those of the original training set (see Table~\ref{tab:contributions}).

\subsection{Average Validation}\label{sec:Average_validation}

To gauge the model's average efficacy, we compared the predicted and true classifications for objects with \(r \leq 21.9\) mag, corresponding to the median 50\% completeness threshold of J-PLUS for compact sources across 1,642 tiles. In Figure~\ref{fig:ROC_curves}, we show the normalized counts for each predicted class in pairs and their respective Receiver Operating Characteristic (ROC) curves. Each pair shows the statistics of True and False positives (TPs and FPs, respectively), therefore showcasing the ability of the model to avoid confusion between species. For example, the Galaxy vs. QSO pair shows how many real galaxies were classified as such (TPs) and how many as QSOs (FPs), while the QSO vs. Galaxy shows the same statistics, but for actual QSOs that are classified as such or as a galaxy.

An ideal classification would yield an area under the curve (AUC) of 1, with no intermixing between classes in the count histograms, e.g. objects with \(y_{\mathrm{true}} = \text{Galaxy}\) would exhibit a probability \(P(\text{class} = \text{Galaxy}) = 1\), while probabilities for other classes would be zero. Our model's results are nearly perfect, with all ROC AUC values exceeding 0.99 for all six possible inter-class combinations. The results also seem to be symmetric between switching classes, with very similar ROC AUC values, indicating that the rate of FPs and False Negatives (FNs) is close to one. Asymmetric results are not desirable, since they reflect biases in the model's prediction towards specific classes. For contrast, \bannjos{} shows its ability to recover well the ratios between classes in all possible combinations, even though the distribution of the test sample is not balanced. The histograms in the upper panels further corroborate the accuracy of \bannjos, displaying nearly all objects congregating at the extremes, indicating \(P(\text{class} = x) \sim 1\) for the evaluated predicted class and \(P(\text{class} = y) \sim 0\) for the other class. As anticipated, the pairs exhibiting the lowest accuracy are Galaxy versus QSO and vice versa, with ROC AUCs of 0.993 and 0.995, respectively, due to the actual dual nature of these sources.

\subsection{Signal-to-Noise Dependence}\label{sec:SN_dependence}

\begin{figure*}
\begin{center}
\includegraphics[width=\linewidth]{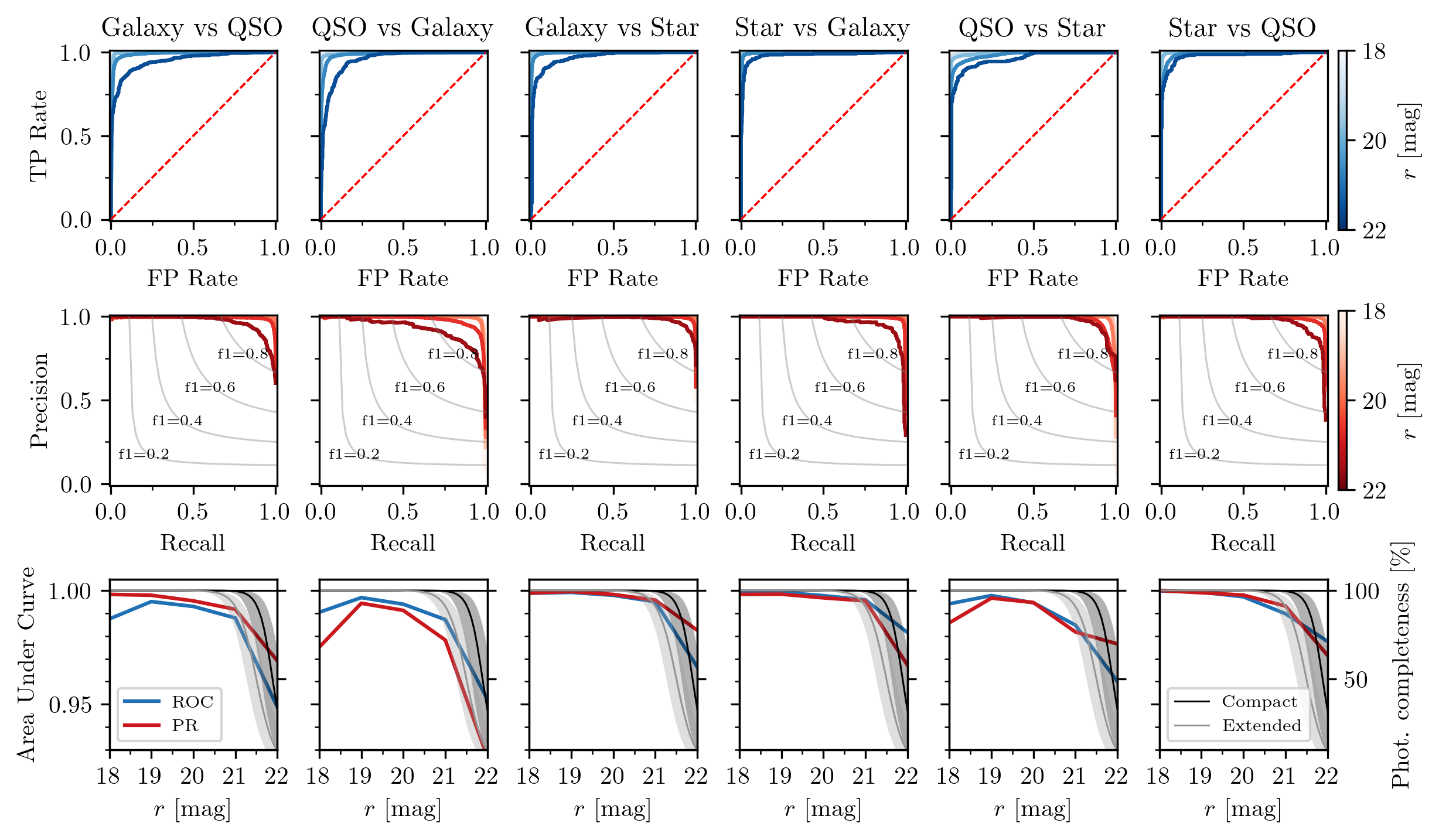}
\caption{Dependence of model performance on the signal-to-noise ratio of the sources. The top panels present the Receiver Operating Characteristic (ROC) curves for objects within different magnitude bins, shaded in varying tones of blue. Magnitude bins range from $r = 17.5$ to $r=22.5$ mag in 0.5 mag increments (10 bins). The middle panels exhibit the Precision-Recall (PR) curves, illustrating the trade-off between sample purity and completeness, shaded in different red tones according to magnitude bin. Both quantities are related by $f1$, the harmonic mean of precision and recall, for which several values are plotted. The bottom panels summarize the results, depicting the evolution of the area under the curve (AUC) for both ROC and PR curves relative to the central magnitude of each bin. Blue and red colors denote the AUCs for ROC and PR, respectively. The median photometric completeness for compact and extended sources within J-PLUS is shown by dark and light gray curves, respectively, with surrounding gray shaded areas indicating the 16th and 84th percentiles of completeness. Classification remains near-perfect up to $r\sim21$ mag (AUC $\gtrsim 0.99$ for both curves), thereafter beginning to diminish, aligning with J-PLUS's limiting magnitude and the ensuing loss of information in some bands. However, even under the most challenging conditions, such as distinguishing QSOs from galaxies or stars, ROC and PR AUC values maintain levels of $\sim$0.98 and $\sim$0.93 at $r = 22$ mag, respectively, showcasing outstanding model performance. Notably, \bannjos{} exhibits a propensity for mistaking QSOs for galaxies and vice versa at magnitudes brighter than $r \sim 18$ mag, possibly due to active galaxies within the test (and training) sample being variably classified as galaxies or QSOs based on the specific spectroscopic survey.}
\label{fig:ROC_curves_permagbin}
\end{center}
\end{figure*}

The predictive capability of \bannjos{} is expected to be influenced by the quality of the data and its signal-to-noise ratio (S/N). We studied the potential impact of lower S/Ns by evaluating ROC curves across various magnitude bins, as shown in Figure~\ref{fig:ROC_curves_permagbin}. This figure also displays Precision-Recall (PR) curves, which balance the purity and completeness of the model's classifications. Ideally, predictions should be both 100\% pure and complete (full Recall). A decline in predictive power results in a compromise between these metrics. As anticipated, the S/N ratio of source measurements in \(\mathbf{x}\) markedly affects prediction quality, with dimmer objects proving more challenging to classify accurately. Deeper magnitude bins, indicated by darker shades, exhibit progressively deviating ROC curves from the ideal upper-left corner as magnitudes increase. This reduction in accuracy is mirrored in the PR curves, with fainter magnitudes resulting in a more obvious trade-off between purity and completeness. The third row in Figure~\ref{fig:ROC_curves_permagbin} illustrates the AUC for both ROC and PR curves as a function of magnitude, alongside the median completeness for both compact and extended J-PLUS sources, depicted in light and darker gray, respectively (do not confuse with the Recall of the classification). The model's classification remains nearly flawless up to $r \sim 21$ mag (AUC $\gtrsim 0.99$ for both curves), before experiencing a decline, coinciding roughly with J-PLUS's limiting magnitude due to diminished S/N and information loss in certain bands. Despite these challenges, even in the least favorable scenarios, such as differentiating QSOs from galaxies or stars, the model maintains high ROC and PR AUC values ($\sim$0.98 and $\sim$0.93 at $r = 22$ mag, respectively), underscoring its exceptional performance. Also noteworthy is the increased confusion between QSOs and galaxies at magnitudes brighter than $r \sim 18$ mag, likely stemming from the presence of active galaxies in both the test and training samples, classified differently across surveys.

\begin{figure*}
\begin{center}
\includegraphics[width=\linewidth]{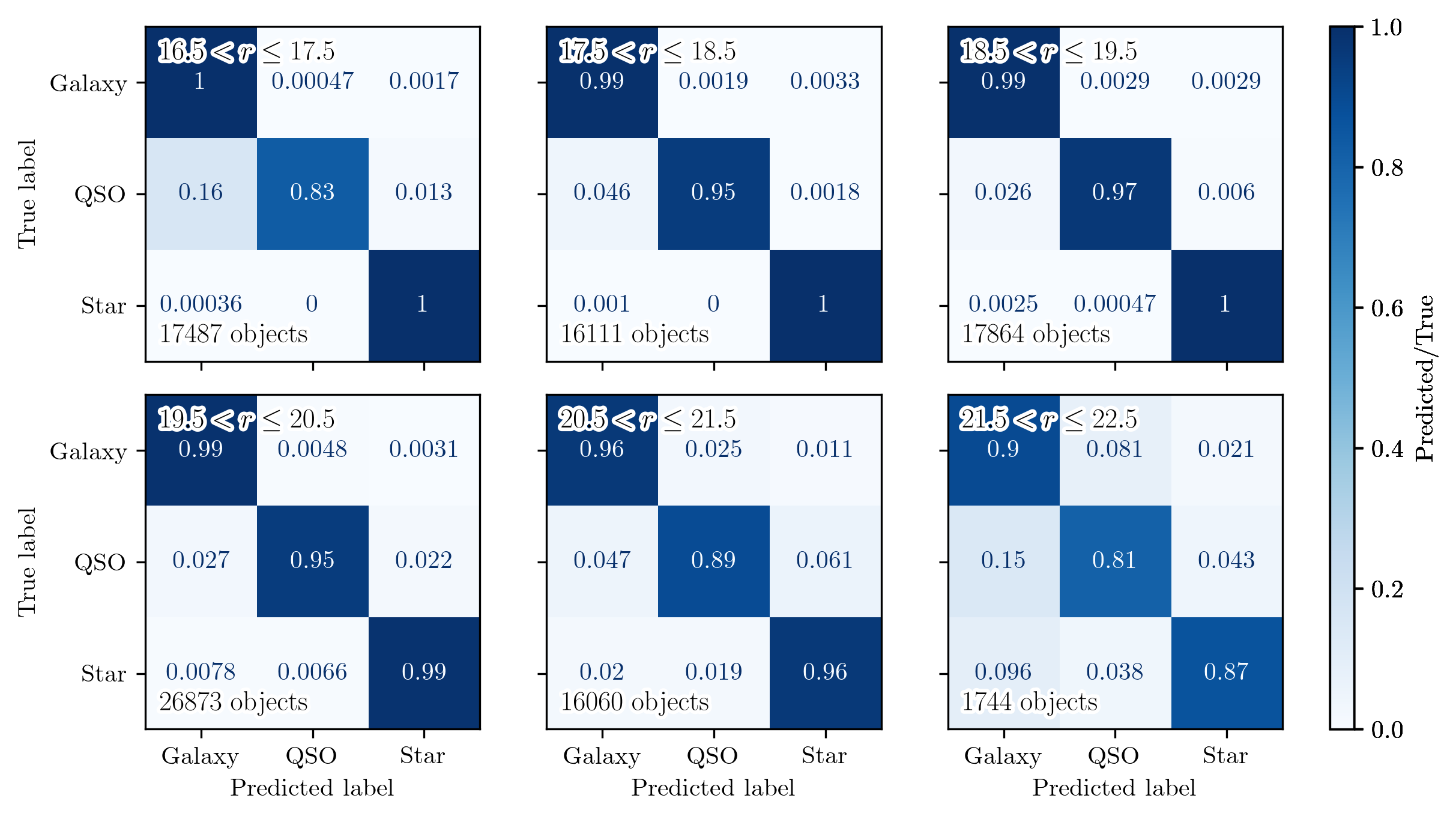}
\caption{Confusion matrices depicting the correlation between the true classes and the predicted classes for objects in the test sample across different magnitude bins. These bins range from $r = 16.5$ to $r = 22.5$ mag in 1 mag increments, totaling six bins. The proportion of objects within each bin relative to the true class is indicated by varying shades of blue, with values represented by real numbers within the interval [0, 1]. Additionally, the confusion matrices include the count of objects contributing to each calculation.}
\label{fig:matrix_permagbin}
\end{center}
\end{figure*}

Further insight into the model's classification performance across different brightness levels is provided by Figure~\ref{fig:matrix_permagbin}, which illustrates the confusion matrices between $y_\mathrm{true}$ and $y_\mathrm{pred}$ for varying magnitude ranges. Overall, the model achieves an average accuracy of approximately 95\% for objects with $r \sim 21.5$ mag. However, it faces challenges in accurately classifying QSOs at fainter magnitudes, a foreseeable issue since QSOs are morphologically similar to stars and their color differentiation becomes less reliable at $r \ge 21$ mag due to the lower S/N ratios. Nevertheless, 81\% of QSOs were correctly identified at $r \ge 21.5$ mag, a good result given that J-PLUS's nominal limiting magnitude is around $r \sim 21.9$ mag. Additionally, since the sample is predominantly composed of stars and galaxies, the average error rate at the faint end remains below 12-13\%.

Similar to observations from Figure~\ref{fig:ROC_curves_permagbin}, there is a noticeable increase in the misclassification of QSOs as galaxies at magnitudes brighter than $r \sim 18$ mag. This trend is attributed to the inclusion in both the test and training samples of active galaxies, which exhibit similar characteristics but are categorized differently as either galaxies or QSOs, depending on the spectroscopic survey in question.

\subsection{Position Dependence}\label{sec:Position_dependence}

\begin{figure*}
\begin{center}
\includegraphics[width=\linewidth]{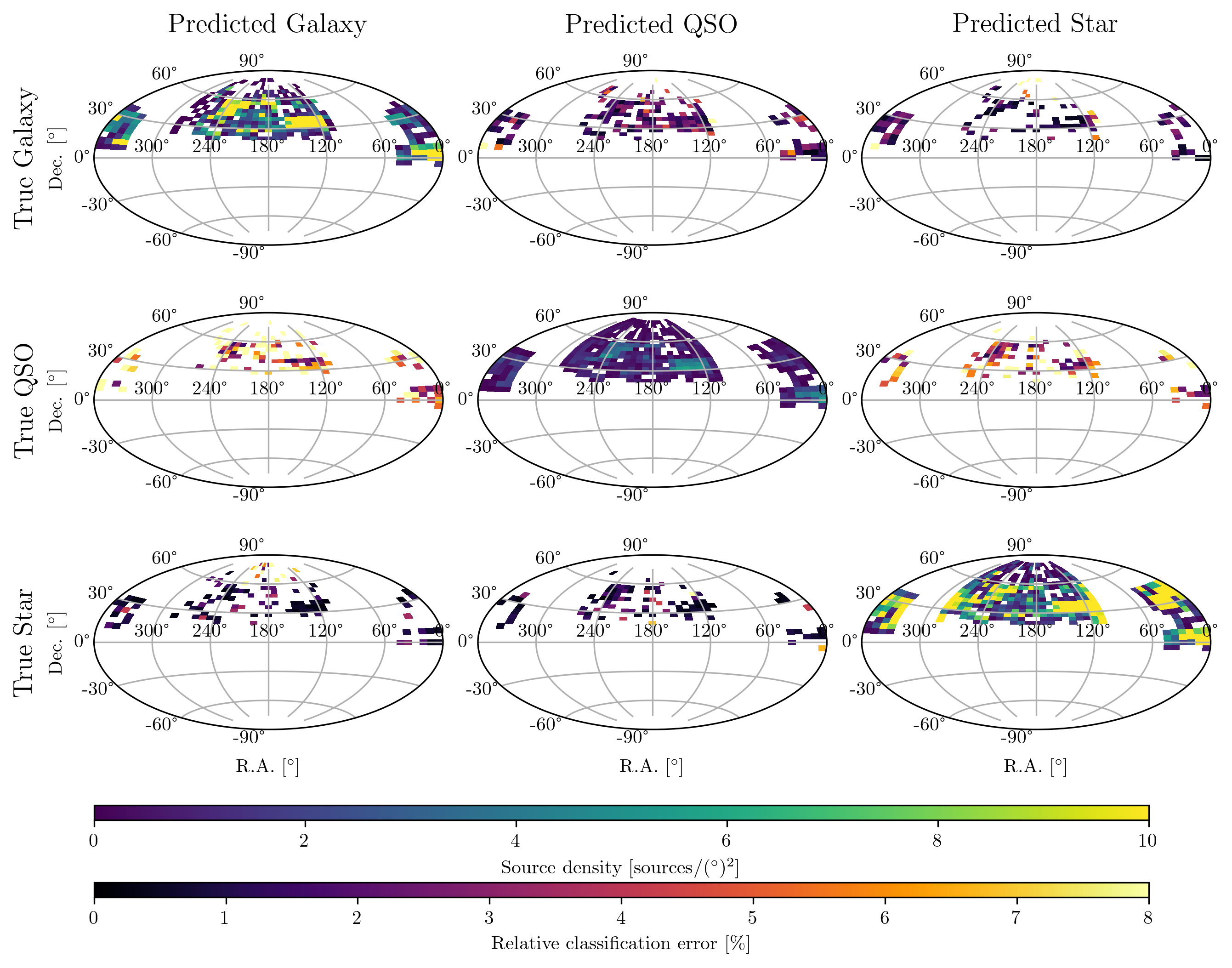}
\caption{Sky maps illustrating the surface density and expected contamination ratios for each class from the test sample. Panels are arranged as in a confusion matrix, with true classes, \(y_{\mathrm{true}}\), on the vertical axis and predicted classes, \(y_{\mathrm{pred}}\), on the horizontal axis. Diagonal panels display the sky surface density of sources classified into each category—galaxies, QSOs, and stars—uniformly scaled in color. The off-diagonal panels exhibit the relative classification error for given true and predicted classes, calculated as the number of misclassified sources divided by the total number of objects in the true class. The off-diagonal panels share a distinct color scale from the diagonal ones. No apparent trends emerge based on sky position.}
\label{fig:classification_error_maps_radec}
\end{center}
\end{figure*}

To ensure our model remains unbiased, we deliberately excluded all positional information; i.e. positions in the sky, two-dimensional proper motions, etc; during its training phase. Nevertheless, due to the incorporation of dust attenuation data from \citet{schlafly11}, some positional data (albeit significantly degraded) may have influenced the model. To identify and mitigate potential biases and to study the presence of contaminants from other classes, we analyzed the prediction errors relative to position and class. This is shown in Figure~\ref{fig:classification_error_maps_radec}, where the diagonal panels show the object surface density for each class in the test sample, and the off-diagonal panels the relative classification error for each possible combination between the three classes, i.e. the percentage of pollutants from each other class. Here, errors are calculated as the number of incorrectly classified objects divided by the total in the corresponding true class for each spatial bin, approximately $3.5^\circ$ on each side.

\begin{figure*}
\begin{center}
\includegraphics[width=\linewidth]{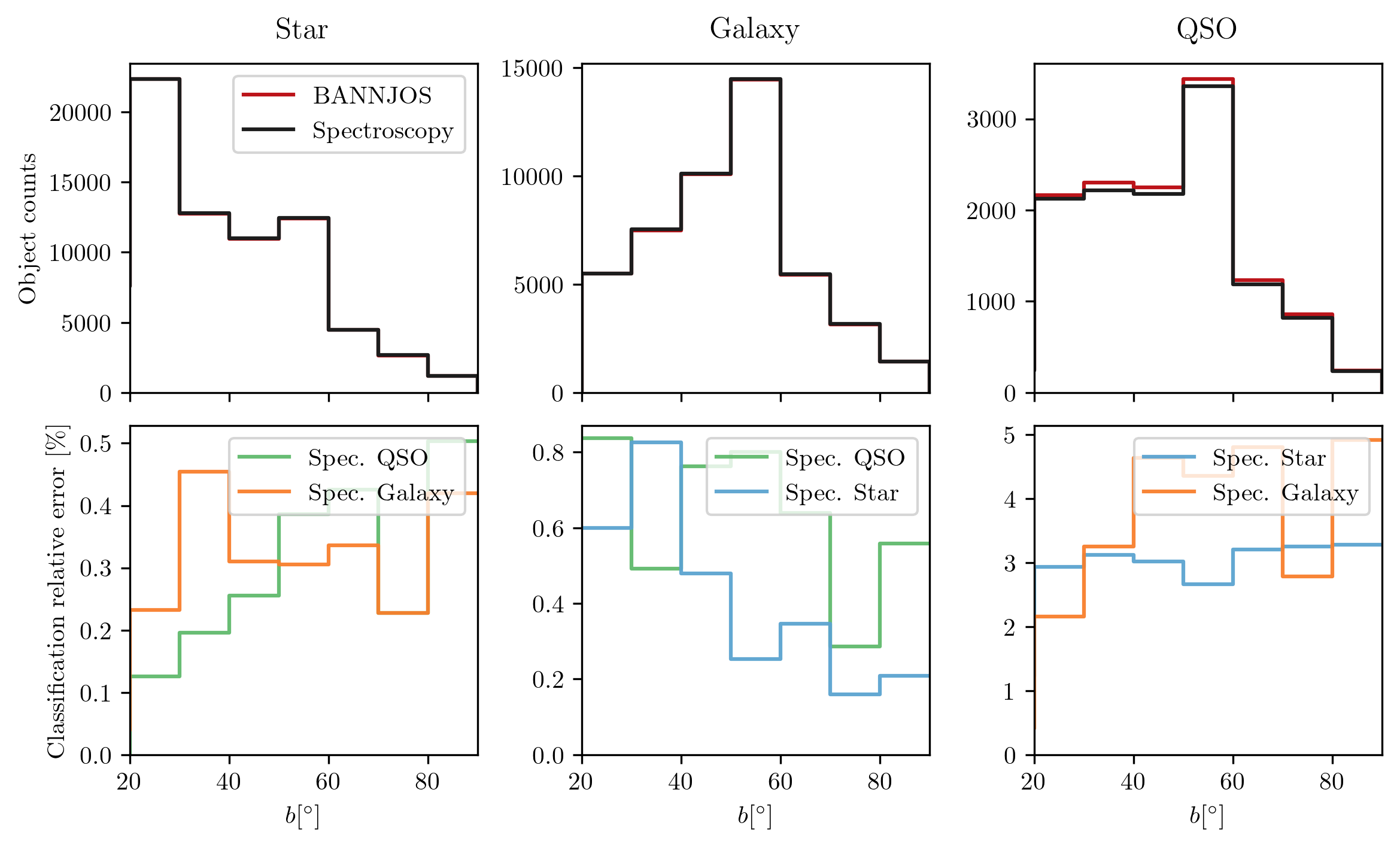}
\caption{Number counts and contamination ratios as function of Galactic latitude, \(b\). {\it Top panels:} Object counts for each class; Star, Galaxy QSO, from left to right; of object classified spectroscopically (black) and by \bannjos{} (red). {\it Bottom panels:} Contamination ratios (false positives) in each class predicted by \bannjos{}. The error is computed as the number of objects incorrectly assigned to each class by \bannjos (one of the other two potential classes), divided by the number of objects of the spectroscopic class, i.e. Star, Galaxy QSO, from left to right. Bins are computed in $10^\circ$ increments from $20^\circ$ to $90^\circ$ for each class. The bin [$10^\circ$,$20^\circ$) is excluded from the analysis due to its very low number of galaxies (47).}
\label{fig:classification_error_rate_galat}
\end{center}
\end{figure*}

Our analysis reveals no significant correlation between source positions and classification error for any of the six class combinations, with error variations across different sky regions consistent with a normal distribution. This includes areas near the Galactic plane, where we might expect an increased prediction of stars if the model were biased by position. The existence of areas with higher surface density, clearly visible in the diagonal panels, is also noteworthy. These correspond to areas that were covered by more surveys or to deeper magnitudes (see Figure~\ref{fig:Aitoff_training}), hence the higher number of observed sources. This inhomogeneity in the training list is the key reason why positional information must not be passed to the model. Otherwise it could learn the specific classification from each spectroscopic survey within their footprints, without generalizing the solution to the entire sky.

We further examined potential Galactic plane proximity effects by assessing the relative error solely as a function of Galactic latitude, \(b\), and class. In Figure~\ref{fig:classification_error_rate_galat}, histograms for each predicted class display the number counts and the average classification error across $10^\circ$ intervals from $20^\circ$ to $90^\circ$. Similar to Figure~\ref{fig:classification_error_maps_radec}, errors are defined as the quantity of objects incorrectly assigned to each class, FPs, ordered by Galactic latitude \(b\). Mild trends emerge here; notably, the misclassification rate of QSOs as stars increases with \(b\), suggesting the model does not entirely compensate for the declining stellar density relative to QSOs. We also notice an increase in wrongly predicted galaxies that are spectroscopically identified as stars at low Galactic latitudes, consistent with higher stellar densities. Conversely, QSO predictions remain relatively stable across latitudes, indicating consistent contamination levels in QSO-predicted samples across all Galactic latitudes. Nonetheless, it's crucial to highlight the scale of demonstrated errors, with the highest error rates being approximately 5\% for predicted QSOs and only 0.8\% and 0.5\% for the galaxies and stars categories, respectively.

\section{Refining Your Selection}\label{sec:Refining_selection}

\begin{figure}
\begin{center}
\includegraphics[width=\linewidth]{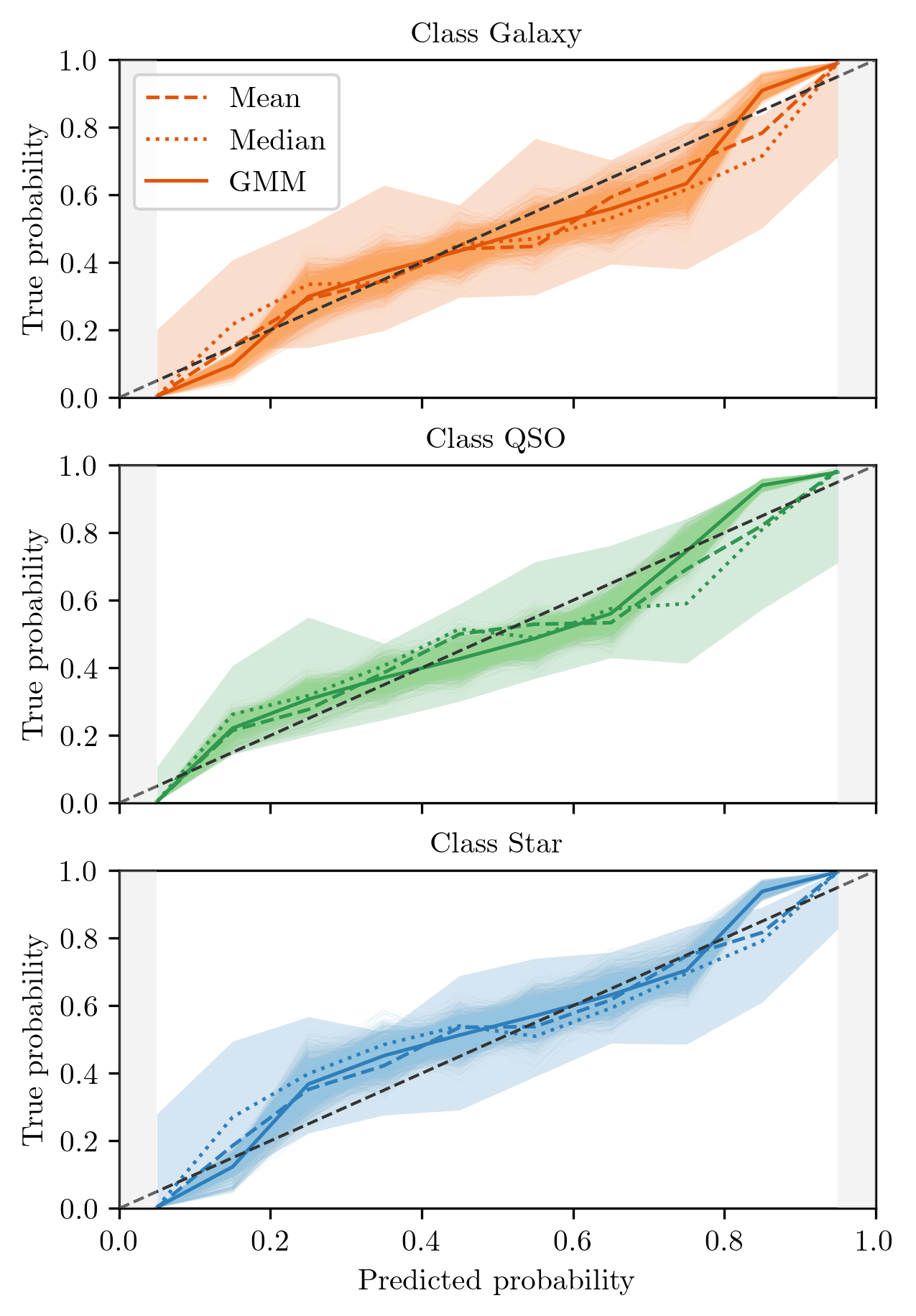}
\caption{Predicted versus true probability for the selection based on different thresholds for each class: Galaxy (top panel), QSO (middle panel), and Star (bottom panel). For each class, the true probability is derived as the amount of objects with confirmed spectroscopic class divided by the total number of objects in each probability interval. The probability intervals are measured on the predicted probability from 0 to 1 in steps of 0.1 (10 in total). The predicted probability is the probability derived with \bannjos{} using three different methods: The mean probability, the median probability, and the mean probability from the reconstructed PDF using the GMM model. The light shaded areas show the true probability of the sample if the 16th or the 84th percentile is used to select the sample. The uncertainties from the GMM model are shown by darker shaded areas and are computed by sampling the model 2000 times. The diagonal dashed line shows the one-to-one relation. Due to the binning used to create the figure, there is only data in the [0.05, 0.95] range.}
\label{fig:predicted_vs_true_probability}
\end{center}
\end{figure}

\begin{figure}
\begin{center}
\includegraphics[width=\linewidth]{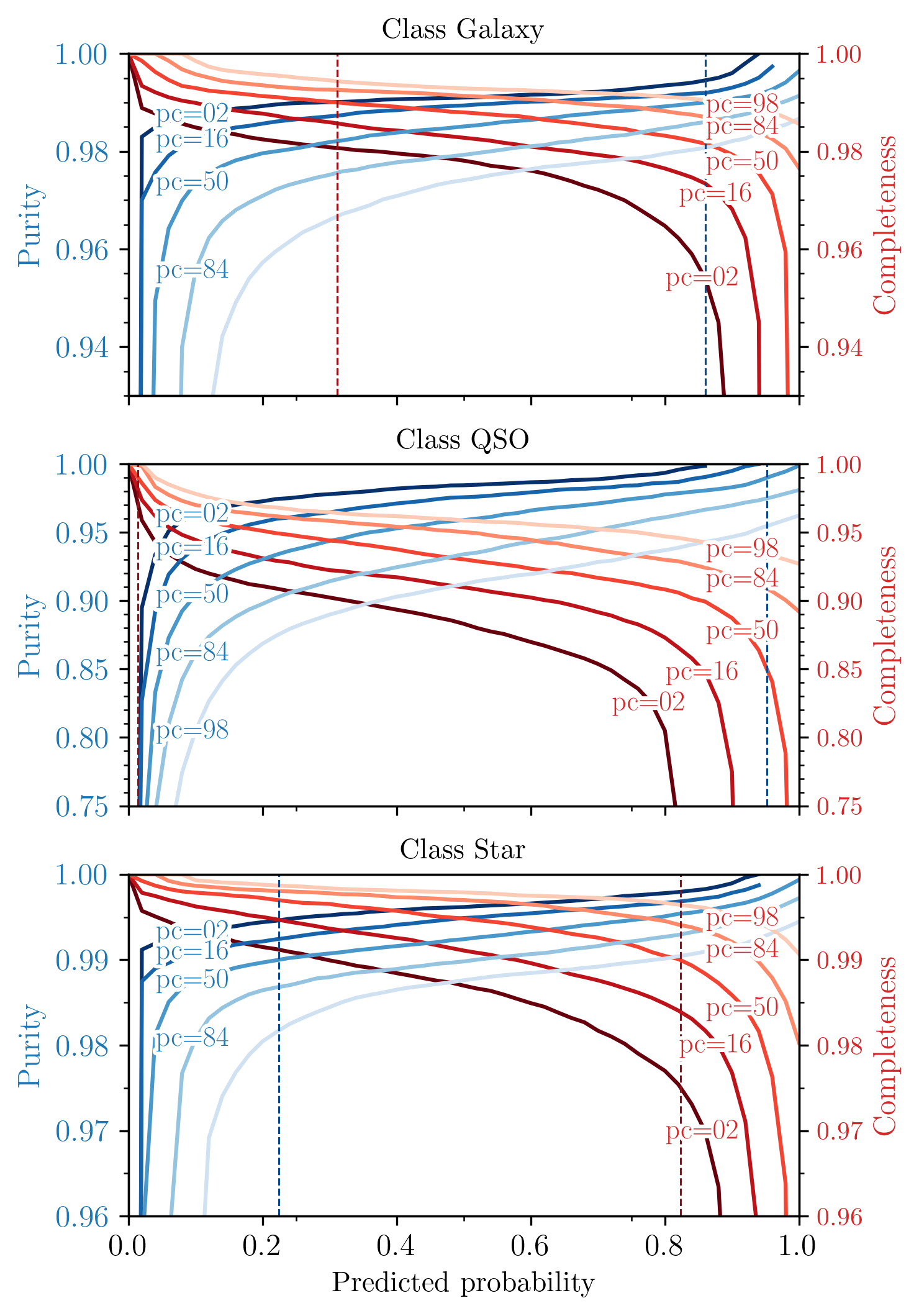}
\caption{Purity versus completeness for the selection based on different thresholds for each class: Galaxy (top panel), QSO (middle panel), and Star (bottom panel). The curves represent purity (blue) and completeness (red) of the sample as functions of the chosen threshold and percentile used to select the class. Vertical dashed lines in corresponding colors indicate the probability thresholds required to achieve 99\% purity (blue) and 99\% completeness (red), respectively.}
\label{fig:completeness_purity}
\end{center}
\end{figure}

\begin{figure*}
\begin{center}
\includegraphics[width=\linewidth]{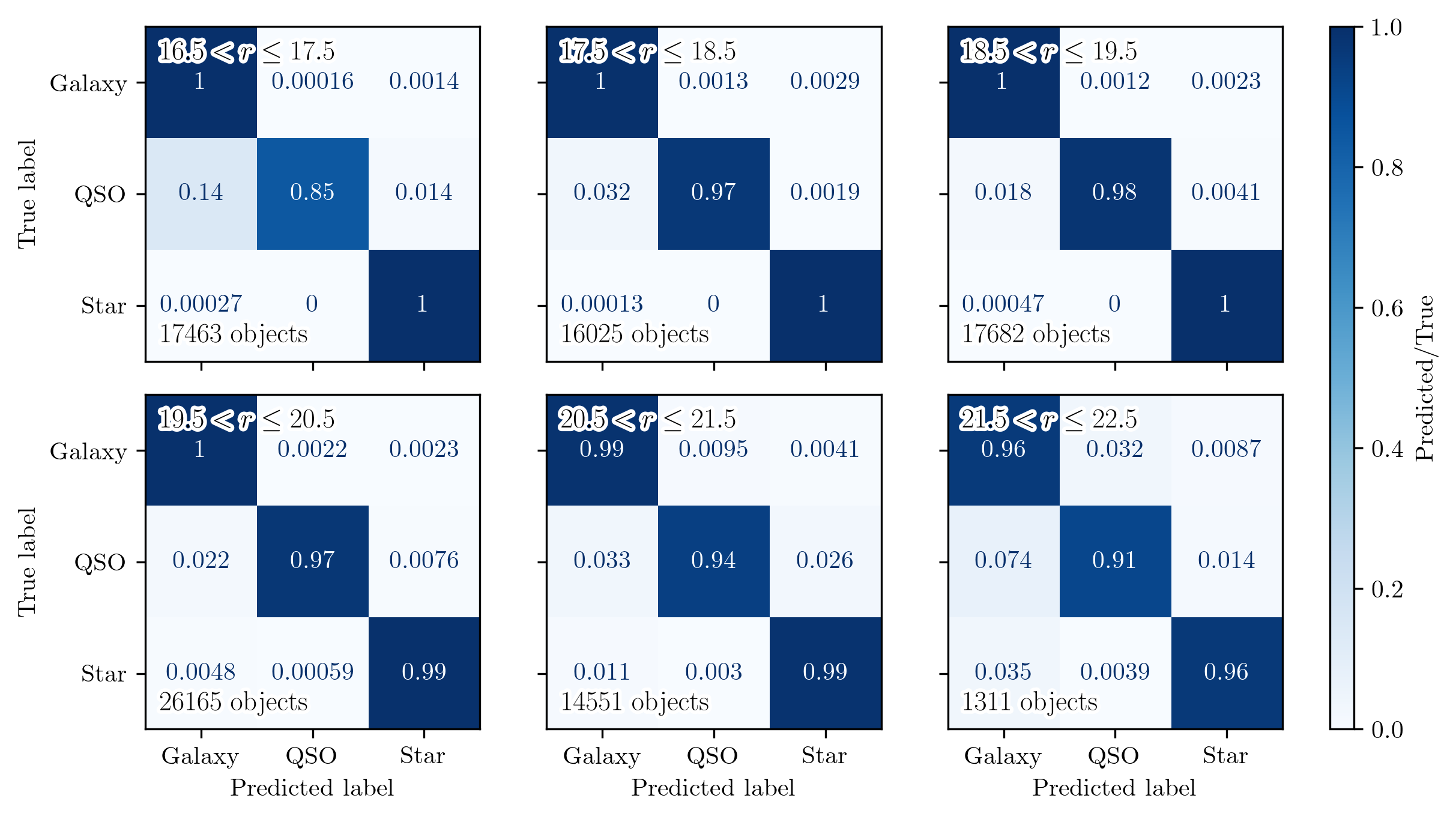}
\caption{Confusion Matrices between the true class and the predicted class for objects in the test sample for different magnitude bins selected with the $2\sigma$ criteria. Markers and symbols coincide with those in Figure~\ref{fig:matrix_permagbin}. The accuracy of the classification is greatly improved when using the predicted uncertainties to select high confidence sources.}
\label{fig:matrix_permagbin_hq}
\end{center}
\end{figure*}

Throughout the paper, we have conventionally assumed that the object's class corresponds to the one with the highest median probability, i.e., \(class = \max \left[ PC_{\text{Galaxy}}(50), PC_{\text{QSO}}(50), PC_{\text{Star}}(50)\right]\). This approach, employed in Sections~\ref{sec:Model_Selection} and~\ref{sec:Validation_model}, does not account for prediction uncertainties. Leveraging the complete PDF for the three classes enables more refined object selection. For instance, applying specific probability thresholds can enhance the purity of the selected sample. Moreover, apart from the median probability, \bannjos{} provides the mean probability and the PDF compressed through the GMM model. These metrics can be used to select classes at any given threshold.

If the model works well, the classification success ratio (true probability) should match the probability used to select the objects (predicted probability). Specifically, the proportion of objects truly belonging to a selected class should correspond to their assigned probability of class membership. This is shown in Figure~\ref{fig:predicted_vs_true_probability}, where the probability predicted by \bannjos{} using three different statistics is compared to the true probability for each class across ten probability bins ranging from 0 to 1. The true probability for each class and probability bin is calculated as the ratio of the number of objects correctly classified according to their \bannjos{} predicted probability and their actual spectroscopic class to the total number of objects with that particular predicted probability.

The three statistical measures yield similar results, with higher predicted probability thresholds correlating to increased true probabilities. The difference between the predicted and true probabilities remains consistently below $\sim 0.1$ across all cases, indicating a good level of accuracy. However, there are some departures from the one-to-one relation, especially when using the median of the probability, that we should point out. For example, \bannjos{} appears to be slightly underconfident when classifying stars with low probability $0.1 \lesssim PC_{\text{Star}}(50) \lesssim 0.5$, and slightly overconfident when classifying galaxies at $0.5 \lesssim PC_{\text{Galaxy}}(50) \lesssim 0.9$. This might result in purer and more contaminated samples of stars and galaxies, depending on the used probability threshold for the median. In contrast, the mean and the reconstructed mean from the GMM model appear to adhere more closely to a one-to-one relation. Figure~\ref{fig:predicted_vs_true_probability} also highlights the probabilistic nature of our findings, with shaded areas representing the sample purity using the 16th and 84th percentiles of the PDF for source selection and the confidence regions derived from the GMM model.

We further investigate the effect of the selection threshold in Figure~\ref{fig:completeness_purity}, which illustrates the impact of varying selection thresholds on the 2nd, 16th, 50th, 84th, and 98th percentiles of the cumulative PDFs, \(x\), for each class: \(PC_{\text{Galaxy}}(x)\), \(PC_{\text{QSO}}(x)\), \(PC_{\text{Star}}(x)\). These curves, similar to the PR curves in Figure~\ref{fig:ROC_curves_permagbin}, demonstrate the purity-completeness trade-off based on the selection threshold.

Increasing the probability threshold typically yields purer samples but reduces completeness. The chosen percentile for selection significantly affects the outcome. For example, selecting objects based on the 98th percentile being a QSO with \(\geq 0.5\) results in a sample with approximately 87\% purity and 98\% completeness. Opting for the 2nd percentile increases purity to roughly 98\% at the expense of dropping completeness to about 91\%. Setting higher thresholds further improves sample purity but invariably reduces completeness. Achieving over 99\% purity in QSO selection is possible by selecting sources whose 2nd percentile exceeds 0.8. Conditions for galaxies and stars are more lenient, with the model classifying them with greater ease. For instance, a purity above 99.5\% is achievable for stars with the 2nd percentile of their PDF exceeding 33\%.

Combining probabilities for the three classes enables the creation of purer samples. Typically, high purity and substantial completeness are attainable by setting the 16th and 84th percentiles above or below the random classification probability (1/3 in this case). By combining these \(1\sigma\) confidence intervals, we can define the object's \(1\sigma\) class as follows:

\begin{itemize}
    \item Star \(\iff PC_{\text{Galaxy}}(84) < 1/3\) \& \(PC_{\text{QSO}}(84) < 1/3\) \& \(PC_{\text{Star}}(16) > 1/3\)
    \item Galaxy \(\iff PC_{\text{Star}}(84) < 1/3\) \& \(PC_{\text{QSO}}(84) < 1/3\) \& \(PC_{\text{Galaxy}}(16) > 1/3\)
    \item QSO \(\iff PC_{\text{Galaxy}}(84) < 1/3\) \& \(PC_{\text{Star}}(84) < 1/3\) \& \(PC_{\text{QSO}}(16) > 1/3\)
\end{itemize}

Adopting the stricter \(2\sigma\) confidence intervals, that is, the 2nd and 98th percentiles instead of the 16th and 84th, yields even purer but less complete samples. 

In Figure~\ref{fig:matrix_permagbin_hq}, confusion matrices between \(y_{\text{true}}\) and \(y_{\text{pred}}\) are presented for sources selected using the \(2\sigma\) criteria for different magnitude ranges. Comparison with Figure~\ref{fig:matrix_permagbin} reveals enhanced sample purity, particularly for QSO classification, which achieves 91\% accuracy in the \(21.5 < r \leq 22.5\) mag range and 96\% for the other classes. However, this precision is gained at the expense of excluding low-confidence classified sources, evidenced by the reduced count of objects in the same magnitude bin (only 1\,311 from 1\,744 objects). The persistent galaxy-QSO confusion at \(r \gtrsim 18\) mag corroborates that these objects are confidently classified by \bannjos{}, suggesting that variation in classification stems from the disparate spectroscopic surveys constituting the training and test sets. The drop from 15\% to 7\% in galaxy-QSO misclassifications at fainter magnitudes supports this notion, as most active galaxies at such depths are identified as QSOs by the employed surveys.

The three selection strategies discussed serve merely as examples of the possible approaches. Higher (lower) selection thresholds can produce more pure (more complete) samples. Even more refined object selections can be achieved by exploiting the full covariance matrix and various correlation coefficients between the classes. For instance, active galaxies could be identified as objects with high probabilities of being either a galaxy or a QSO, accompanied by a significant negative correlation between these probabilities, as discussed in Section~\ref{sec:training_sampling}. If very pure samples are required, the correlation coefficients between probabilities can be useful too, by selecting sources with low uncertainties and little degeneration between species. Lastly, since the full PDF is recoverable from \bannjos's output, users can also opt to use it for weighting the PDFs, finding particular objects, etc.

In Appendix~\ref{Appendix:Selecting_pure_samples}, we provide examples on how to query \bannjos{} data to obtain pure samples of specific objects. In particular, we demonstrate how to select a pure sample of QSOs that could be candidates for spectroscopic follow up.

\section{Comparisons with other classifiers}\label{sec:Comparison_previous}

At the moment of writing this paper, there are three other classifications available for J-PLUS. All of them are deterministic, meaning that they provide a single value for the probability of each object belonging to a certain class. The most important difference between them lies in the basis of their classification, morphological or nature-based, and thus in the number of classes they manage. In the following subsections, we compare \bannjos's predictions to these previous works by using the highest median as the assigned class.

\subsection{two-classes classifiers}\label{sec:Comparison_simple}

\begin{figure*}
\begin{center}
\includegraphics[width=\linewidth]{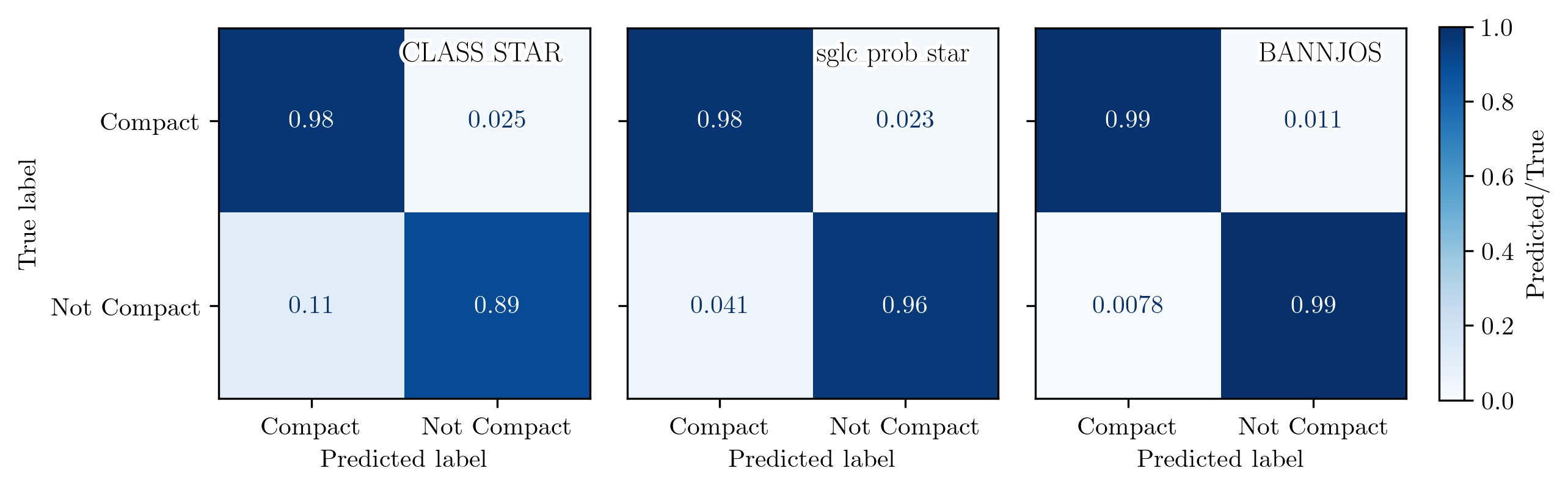}
\caption{Confusion matrices comparing the true class (spectroscopic) against the predicted class for objects in the test sample up to $r = 21.9$ mag, corresponding to the median 50\% completeness threshold of J-PLUS for compact sources across 1,642 tiles, for three different classifiers. These matrices are based on the assumption that QSOs and stars are compact sources, while galaxies are extended. A source is considered compact if its corresponding \texttt{CLASS\_STAR}, \texttt{sglc\_prob\_star} scores, or its median $PC_{\rm Star | QSO}(50)$ exceeds 0.5. Labels coincide with those of Figure~\ref{fig:matrix_permagbin}.}
\label{fig:matrices_class_star_bannjos}
\end{center}
\end{figure*}

\begin{figure*}
\begin{center}
\includegraphics[width=\linewidth]{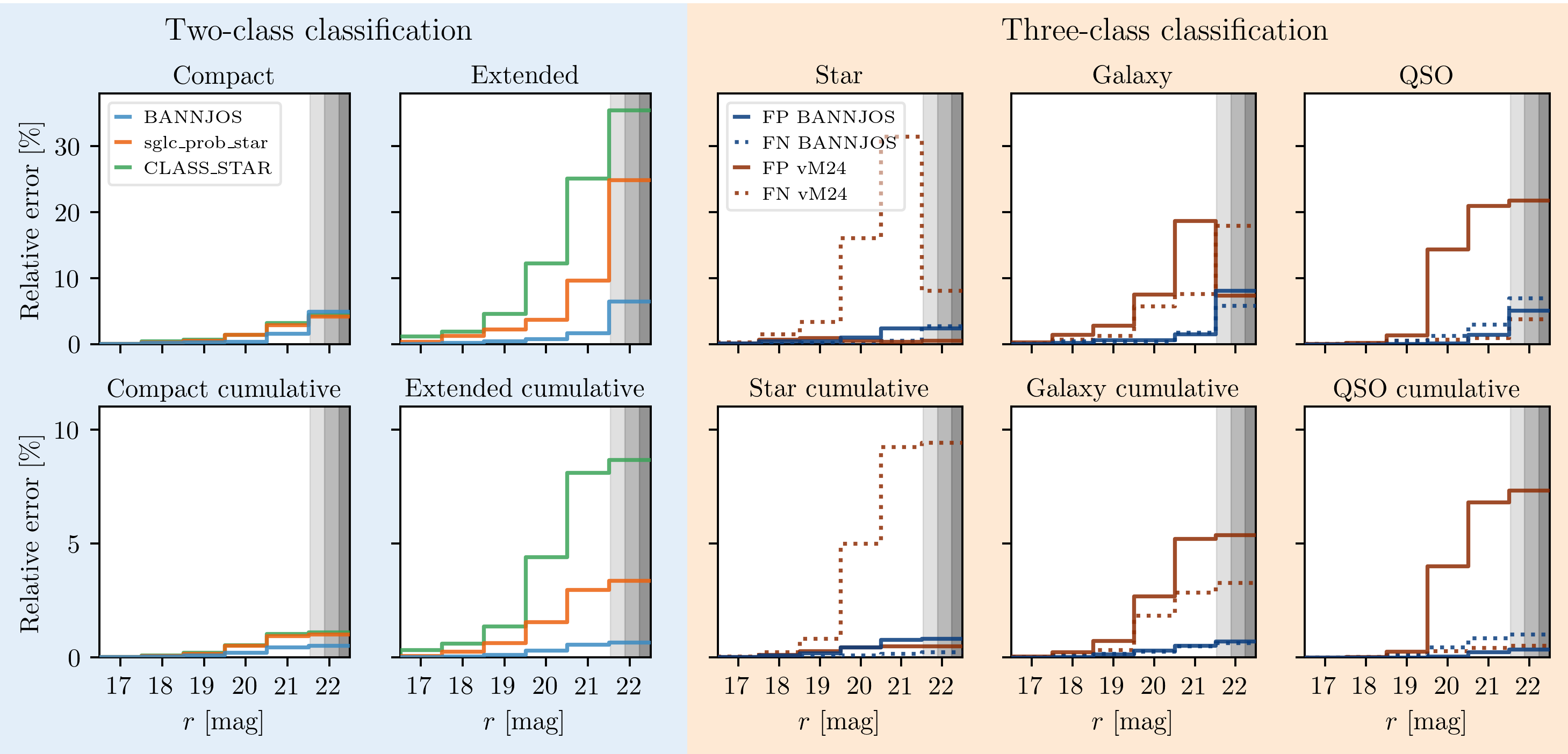}
\caption{Classification error rates per magnitude bin for all evaluated classifiers. The magnitude bins coincide with those in Figure~\ref{fig:ROC_curves_permagbin}, and error rates are defined as the ratio of objects incorrectly classified as False Positives (FPs) or False Negatives (FNs) to the total number of objects within each bin. \textit{Left sub-figure (Two-class classification, light blue):} The error rates for compact objects, specifically galaxies objects misclassified as stars or QSOs (left) and vice versa (right), are presented. Each classifier is denoted by a unique color: \bannjos{} in blue, \texttt{sglc\_prob\_star} in orange, and \texttt{CLASS\_STAR} in green. An object is considered compact if its corresponding \texttt{CLASS\_STAR} or \texttt{sglc\_prob\_star} score exceeds 0.5, or its median $PC_{\rm Star | QSO}(50)$ is greater than $1/3$. The statistics are obtained using the test sample of 136,570 objects. \textit{Right sub-figure (light orange):} Shows the classification error rate for the three classes available in \bannjos{} (dark blue) and \citet{vonmarttens24} (vM24, dark orange). FPs are depicted with solid lines, while FNs are shown with dotted lines in the respective colors. In this case, the error classification ratios are obtained with another test sample of 52,105 sources never seen during the training phases of the two models. The distribution of errors over cumulative magnitude bins is shown in the corresponding bottom panels. The median 90\%, 50\%, and 10\% completeness levels for J-PLUS compact sources are indicated by vertical gray shaded areas, with the darkest shade representing the 10\% level. \bannjos{} surpasses all other classifiers by a large margin at all magnitude range, both in total classification error and in symmetry between FPs and FNs.}
\label{fig:classification_error_rate_comparison}
\end{center}
\end{figure*}

Two classifications are currently available along J-PLUS tables at CEFCA's archive. The first, \texttt{CLASS\_STAR}, is derived from the \texttt{SExtractor} photometry package. The second, \texttt{sglc\_prob\_star}, introduced by \citet{lopezsanjuan19}, applies Bayesian priors to enhance the classification accuracy beyond that provided by \texttt{SExtractor}. Both classifiers categorize sources into two morphological classes: compact or extended, thereby assigning the probability of a source being star-like. However, this binary scheme complicates direct comparisons with our three-class approach. To address this, we recategorized the true and predicted classes from our test sample ($y_\mathrm{true}$ and $y_\mathrm{pred}$) into these two morphological categories. We classified objects as compact if their \texttt{CLASS\_STAR}, \texttt{sglc\_prob\_star} scores, or their median $PC_{\rm Star | QSO}(50)$ exceeded 0.5. Conversely, objects were deemed extended otherwise. This approach enables direct comparison between our results and those of the pre-existing binary classifiers.

Figure~\ref{fig:matrices_class_star_bannjos} presents confusion matrices for these morphological classification derived from the three classifiers. All three classifiers perform commendably in classifying compact sources, with correct classification rates of approximately 98\%. However, \texttt{CLASS\_STAR} exhibits notable shortcomings in identifying extended sources, correctly classifying only about 89\% as galaxies. The \texttt{sglc\_prob\_star} classifier significantly improves upon this, correctly identifying around 96\% of galaxies. When examining the misclassification of galaxies as compact (star-like) sources, \texttt{sglc\_prob\_star} shows a threefold improvement over \texttt{CLASS\_STAR}. Nevertheless, \bannjos{} demonstrates superior accuracy, significantly reducing the misclassification rate of extended objects compared to \texttt{sglc\_prob\_star}.

The left part of Figure~\ref{fig:classification_error_rate_comparison} (blue) illustrates the classification error rates for the three classifiers across various magnitude bins. Errors for each classifier are quantified as the ratio of misclassified objects to the total number of objects per bin, and are further categorized into False Positives (FPs) and False Negatives (FNs). Since only two classes are involved, the FNs in compact sources correspond to the FPs in extended sources. \bannjos{} consistently outperforms the other classifiers across all bins, maintaining lower error rates. This holds true except possibly for FPs (Compact) at the faintest magnitudes, where the error rates for all three classifiers converge. A notable aspect in this comparison is the significant asymmetry displayed by the \texttt{sglc\_prob\_star} and \texttt{CLASS\_STAR} classifiers between FPs and FNs, indicating a tendency to misclassify QSOs or stars as extended sources as the S/N decreases, with \texttt{CLASS\_STAR} being the most pronounced in this regard.

Adding FPs and FNs yields the total classification error, where \bannjos{} outperforms the other two classifiers by a significant margin. For example, \texttt{CLASS\_STAR} incorrectly classifies approximately 30\% of objects beyond $r > 20.5$ mag, resulting in a cumulative error rate of about 11\% at $r = 22.5$ mag. In contrast, \texttt{sglc\_prob\_star} reduces the error rate to approximately 12-13\% at similar magnitudes, with an accumulated error of about 5\% at $r = 22.5$ mag. \bannjos{}, however, achieves an error rate of merely around 3.5\% in the same range, with a cumulative error below 2\% for all objects. Remarkably, \bannjos{} maintains near-perfect performance for objects brighter than $r = 19.5$ mag, with error rates around 0.3\% and cumulative errors below 0.1\%. While the purity of the classification could potentially be enhanced by adjusting the probability thresholds for the three classifiers, our experiments indicate that the relative differences between the classifiers remained almost unchanged, or favored \bannjos{} even more.

\subsection{Three-classes classifiers}\label{sec:Comparison_three}

Differently from \texttt{CLASS\_STAR} and \texttt{sglc\_prob\_star}, the classification introduced in \citet{vonmarttens24}, hereafter referred to as vM24, bases its results not only on the morphology of the source but also on its colors. This approach, coupled with the utilization of a more sophisticated algorithm, \texttt{XGBoost}, enables the authors to differentiate sources based on their inherent nature, adding the QSO class to their classification. Consequently, their classification encompasses three categories: Galaxy, QSO, and stars; making it a natural counterpart for comparison with our classification.

A new test sample is required for this comparison, one that includes only objects never seen during the training phases of any of the models, i.e., \bannjos{} and \texttt{XGBoost}. This new test sample was composed by cross-matching our own test sample with the training sample used in vM24, retaining only sources present in our test sample but absent in their training sample. From our original test sample of 136,570 objects, we identified 52,105 sources not included in the vM24 training sample. While significantly reduced, this number of objects is still sufficiently large to conduct a comparative analysis. This dataset, composed entirely of sources unfamiliar to both models, allows us to assess their general performance on independent data, thus facilitating a fair comparison between the two. As in previous sections, we assign the class predicted by \bannjos{} as the one with the highest median probability, $\max[PC_{\text{class}}(50)]$. In vM24, the classification is presented similarly to that of \bannjos{}, albeit with a single probability value per class. We assigned the class in vM24 as the one with the highest probability. While simple, this is the only criterion ensuring 100\% completeness in the sample selection.

Following the analysis from the previous section, we proceeded to bin the new test sample according to the $r$-magnitude of its sources. We then computed the relative and cumulative errors as in the left sub-figure of Figure~\ref{fig:classification_error_rate_comparison}, i.e., the relative classification error per class for each classifier is defined as the ratio between misclassified objects and the total number of objects, determined by their spectroscopic class, within each magnitude bin. The results are shown in the right part of Figure~\ref{fig:classification_error_rate_comparison} (orange), where upper panels detail the relative FP and FN errors for each classifier by class, and bottom panels present cumulative errors for FPs and FNs.

\bannjos{} significantly outperforms the performance presented in vM24 across all magnitude bins, with accuracies varying from 2 to 10 times better, depending on the class and magnitude bin. The higher error rate of vM24 is clearly visible as a generally higher rate of FPs and FNs in the vM24 results compared to those of \bannjos{}. For instance, the typical total classification error (FPs+FNs) for Galaxies at $r = 21$ mag is around 27\% in vM24, while \bannjos{} reduces this to merely 3\%. These differences diminish at the faintest magnitudes, where \bannjos{} still typically yields classification accuracies twice as higher as those obtained in vM24. To provide some figures, the average total classification error for QSOs in vM24 is approximately 26\% at $r \sim 22$, compared to around 12\% for \bannjos{}, while cumulative errors typically escalate to 8\% at $r \sim 22.5$ in vM24 but remain at $\sim1.5$\% for \bannjos{}.

Asymmetries between FPs and FNs could indicate a bias in the model's predictions towards a specific class. Notably, stars exhibit a very high rate of FNs in vM24, which seems to translate into FPs for galaxies and QSOs, indicating that at magnitudes fainter than $r \sim 18$ mag, the \texttt{XGBoost} used in vM24 tends to misclassify a significant number of stars as galaxies and QSOs. Confusion also seems to occur between the galaxy and QSO classes at the faintest magnitudes. This effect is also observable in \bannjos{}'s results, albeit with a lower error rate and more symmetry between FPs, FNs, and the classes themselves. However, as discussed in Section~\ref{sec:Average_validation}, this is expected and likely caused by the dual nature of active galaxies.

We conducted an additional test using a refined sample, specifically selecting sources classified with over 95\% probability in vM24 in any of the three classes. This criterion reduced the sample size to 37,355 objects (approximately 72\% of the original sample) that are classified with high confidence in vM24. Upon reevaluating both models with this high-probability sample, improvements were noticeable, although \bannjos{} continued to exhibit superior performance across all magnitude bins. In vM24, typical error rates peaked at around 8\% at $r \sim 21$ mag for both stars and galaxies, with QSOs remaining lower at 2\% at the same magnitude. However, the error distribution between these classes remained very asymmetric, with classification errors for stars being almost entirely FNs and for galaxies mostly FPs. The high asymmetry shown between the galaxy and star classes at magnitudes fainter than $r \sim 19$ mag indicates that the vM24 classification systematically misclassifies stars as galaxies at these magnitudes when applying high-probability selection criteria. Interestingly, the numbers of FPs and FNs are consistent for QSOs under this high-probability selection. In contrast, \bannjos{} exhibits its maximum error at the faintest magnitudes, $r = 22.5$ mag, at around 2\% for galaxies and around 1\% for the other classes, demonstrating higher consistency between FPs and FNs across the entire magnitude range. In summary, using this particular high-confidence sample from the vM24 catalog, \bannjos{} still outperforms \texttt{XGBoost} significantly.

The largest difference between the training of the models stems from the addition of \DESI{} data to our training list. The test sample used for this comparison also contains sources classified by \DESI, which could potentially represent and advantage for \bannjos{}. To test whether the presence of these sources in the test sample was beneficial to \bannjos{}, we removed all the \DESI-only-measured sources from the list, leaving a total of 31,124 sources, and repeated the experiment. While \bannjos{}'s errors remained the same or even decreased in some cases, the classification errors in vM24's results increased, particularly in the star and galaxy classes.

All previous tests were performed covering scenarios going from neutral to favorable to vM24. In a last test, instead of selecting objects based on vM24 scores, we repeated the experiment using the $2\sigma$ criteria presented in Section~\ref{sec:Refining_selection}, including sources classified by \DESI. This criterion, which could potentially be favorable to \bannjos, yielded 51,214 sources (approximately 98\% of the original). As expected, the results using this list show a significant improvement in the classification accuracy of \bannjos{} (with cumulative errors well below 1\% for the three classes) while still maintaining a large portion of the sources. In contrast, the \texttt{XGBoost} model used in vM24 yielded slightly worse results than those obtained using the entire sample, potentially due to the marginal reduction in sample size without a corresponding decrease in the number of misclassified objects.

In Appendix~\ref{Appendix:vM24_star}, we offer a comparison between the four classifiers, i.e. \texttt{CLASS\_STAR}, \texttt{sglc\_prob\_star}, vM24, and \bannjos, based on ROC curves for this new test sample of 52,105 sources.

\section{Results and Statistical Validation}\label{sec:Results_final_validation}

In the previous section, Section~\ref{sec:Validation_model}, we validated the classification using the test sample, composed of sources with a spectroscopic classification \textit{never seen} by our model. The test sample represents a valuable asset for validating our classification. However, it is still a subset of the training set, hence the model might have certain advantages when classifying its objects. For instance, the observed properties of sources in both the test and training sets are similar, and the S/N is relatively high for most sources. This scenario could lead to overly optimistic validation results, as the model might merely learn to reproduce the classifications for the types of objects found in the training set.

To further validate the classification performance of \bannjos{}, we compared its predictions with those from previous independent works across the entire J-PLUS DR3 catalog. We employed \bannjos{} to classify all 47.4 million objects in J-PLUS DR3 into stars, QSOs, and galaxies. In the following subsections, we statistically analyze the classification results for the entire J-PLUS dataset.

\subsection{Completeness}\label{sec:Completeness}

\begin{figure}
\begin{center}
\includegraphics[width=\linewidth]{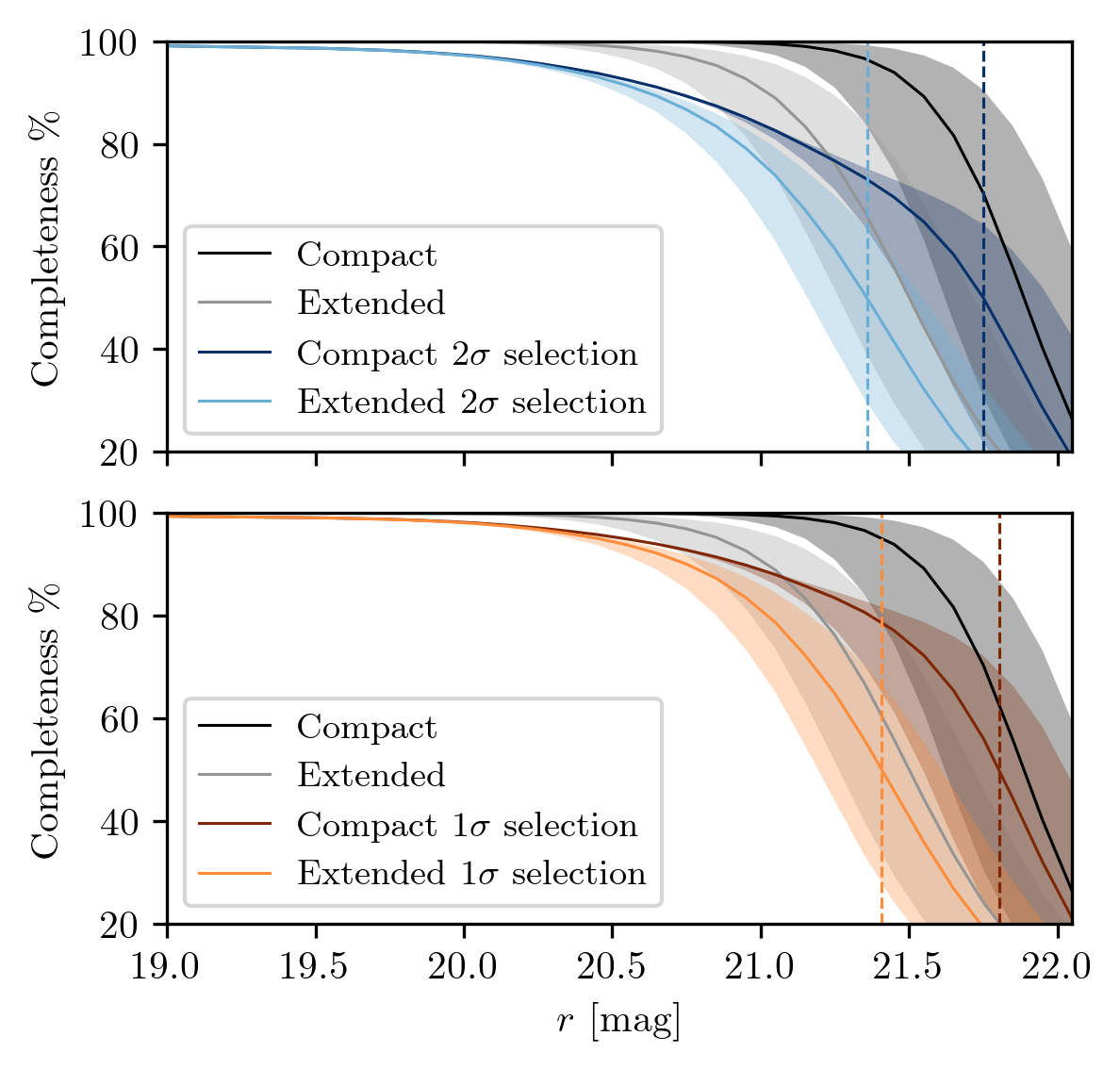}
\caption{Final completeness of the J-PLUS survey using different selection criteria in \bannjos. The black and grey curves represent the median photometric completeness of J-PLUS for compact and extended sources, respectively, measured across its 1,642 tiles in 0.1 mag wide magnitude bins. This is equivalent to the completeness achieved when selecting sources based on \bannjos{}'s maximum median probability. The shaded areas indicate the 16th and 84th percentiles of this completeness. The orange curves and their corresponding shaded areas in the bottom panel illustrate the completeness for compact and extended sources using $1\sigma$ significance in their Probability Density Functions (PDFs). The blue curves in the upper panel represent the completeness using the $2\sigma$ selection criteria (see Section~\ref{sec:Refining_selection}). Vertical dashed lines indicate the 50\% completeness level for each configuration, and are colored correspondingly.}
\label{fig:Final_completeness}
\end{center}
\end{figure}

Initially, we examine the completeness of the classification based on the three basic selection criteria presented in Section~\ref{sec:Refining_selection}. Figure~\ref{fig:Final_completeness} shows the average completeness of J-PLUS as a function of $r$ magnitude for the three criteria: highest median probability, $1\sigma$, and $2\sigma$. The completeness is calculated as the number of sources remaining after classification divided by the total number of sources, then multiplied by the actual photometric completeness of J-PLUS at each specific magnitude. Thus, the curves represent the expected real completeness for compact and extended sources at a given magnitude $r$ when using the \bannjos{} classification with these selection criteria. We avoid detailed calculations for the three individual classes since we only have access to the classes predicted by \bannjos, which could contain misclassified objects, especially at fainter magnitudes.

Using the highest median value of the PDFs ensures no object is rejected, hence the completeness remains equivalent to the photometric completeness of J-PLUS for both compact and extended sources (dark and grey curves, respectively). However, selecting sources based on their confidence intervals leads to the exclusion of uncertain sources. As expected, the number of sources excluded depends on the visual morphology of the sources and the considered magnitude range. In general, the $1\sigma$ selection is less complete than using the maximum median value. Yet, completeness values above 95\% are still achievable at $r \approx 20.3$ mag, and 90\% at $r \approx 21.2$ mag for compact sources. The selection based on $2\sigma$ confidence intervals is more restrictive, placing the same levels of completeness at $r \approx 20.0$ and $r \approx 20.5$ mag, respectively.

\subsection{Number Counts}\label{sec:Number_counts}

\begin{figure}
\centering
\includegraphics[]{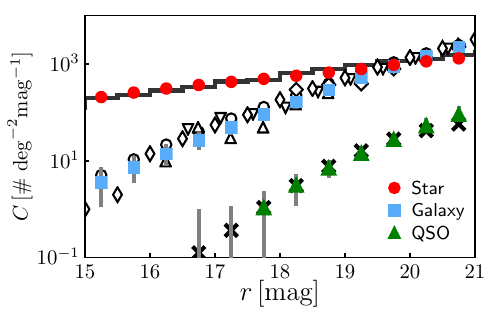}
\includegraphics[]{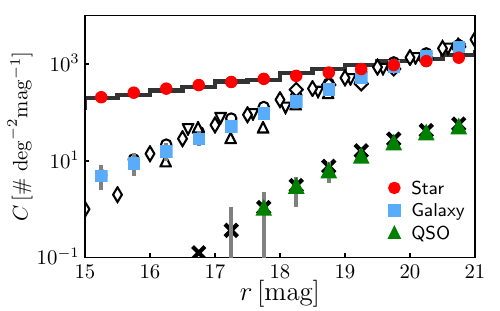}
\caption{\textit{Top panel:} Number counts for stars (red dots), galaxies (blue squares), and QSOs (green triangles) as a function of $r$-band magnitude, estimated as the median of the 1642 J-PLUS DR3 pointings, using the maximum median probability criteria to assign classes. Vertical gray lines show the variance in number counts between pointings. Black solid histograms represent stellar number counts at the median pointing position, estimated using the TRILEGAL model of the Milky Way \citep{trilegal}. Open symbols denote galaxy number counts from literature sources: \citet[][circles]{yasuda01}; \citet[][triangles]{huang01}; \citet[][inverted triangles]{kummel01}; and \citet[][diamonds]{kashikawa04}. Black crosses show QSO number counts as predicted by \citet{palanque2016}. \textit{Bottom panel:} Similar to the top panel, but applying the more restrictive $2\sigma$ selection criteria.}
\label{fig:counts_ps}
\end{figure}

Table~\ref{tab:Final_counts} presents the final numbers of objects classified by \bannjos: 47,463,878, divided into galaxies, QSOs, and stars using the three different criteria presented in Section~\ref{sec:Refining_selection}. As criteria become more restrictive, fewer objects are classified with sufficient confidence. For instance, of the approximately $2.0 \times 10^7$ galaxies observed in J-PLUS and classified with the $\max[PC_{\text{class}}(50)]$ criterion, only about $1.7 \times 10^7$ are classified as such with a $2\sigma$ confidence. This effect is most pronounced for QSOs, where fewer than half of those classified with the $\max[PC_{\text{class}}(50)]$ criterion meet the high confidence threshold of the $2\sigma$ criteria. Stars, however, suffer the least impact from poor S/N; thus, the three classification criteria produce nearly the same number of stars.

\begin{table} 
\caption{Object counts for the three different criteria discussed in Section~\ref{sec:Refining_selection}}
\label{tab:Final_counts}
\centering 
        \begin{tabular}{l c c c}
        \hline\hline\rule{0pt}{3ex} 
        Class   & $\max[PC_{\text{class}}(50)]$  & $1\sigma$  & $2\sigma$ \\ 
        \hline
        Galaxies   & 20,479,247  &   18,466,693   &   17,391,099   \\ 
        QSOs       & 997,436     &   508,440      &   404,010      \\ 
        Stars      & 25,987,195  &   24,712,265   &   24,175,950   \\ 
        Total      & 47,463,878  &   43,687,398   &   41,971,059   \\
        \hline 
\end{tabular}
\end{table}

\bannjos's classification should yield number counts consistent with expectations from models and previous works. We computed the number counts for each object type as a function of $r$ magnitude using the $\max[PC_{\text{class}}(50)]$ criterion, which includes all sources defining the class as the one with the highest median probability. We defined the $r$-band number counts in each J-PLUS pointing as the Probability Density Function (PDF)-weighted histogram normalized by area and magnitude (see Eq. 17 in \citealt{lopezsanjuan19}). The median and dispersion of counts from the 1642 pointings were then computed. We present the J-PLUS DR3 stellar, galaxy, and QSO number counts in Fig.~\ref{fig:counts_ps}. The stellar number counts match the TRILEGAL model \citep{trilegal} predictions for the Milky Way at the median (RA, Dec) of the J-PLUS DR3 tiles, despite a large dispersion reflecting the variation in stellar density across the surveyed area. The galaxy number counts is consistent with literature results. However, QSO number counts exhibit an excess over expectations from \citet{palanque2016} at $r > 20$, with approximately 30\% and 60\% more QSOs at $r = 20.25$ and $20.75$ mag, respectively. Notably, QSOs are 100 times less numerous than galaxies and stars at faint magnitudes, and even a low misclassification rate can significantly overestimate the number of QSOs.

Repeating the analysis with a more stringent $2\sigma$ selection scheme for QSOs (bottom panel), we found that while stellar and galaxy number counts remained unchanged, QSO number counts at faint magnitudes decreased. In this case, the QSO counts are at 90\% of the expectations for sources within $19 < r < 21$ mag. This value is comparable to the predicted 91\% completeness for the $2\sigma$ selection outlined in Section~\ref{sec:Refining_selection}, suggesting that the selection could indeed be of high purity. This underscores the utility of classification PDFs beyond the best solution, illustrating that \bannjos{} can statistically provide accurate densities of stars, galaxies, and QSOs for $r < 21$ mag. 

It is also important to point out that \bannjos{} was trained using a list that is balanced, containing an equal number of objects for each class. The fact that \bannjos{} accurately estimates the number counts for the three classes showcases its ability to distinguish between classes without propagating biases regarding the expected number of objects.

\subsection{Color-Color Diagrams}\label{sec:color-color}

\begin{figure*}
\begin{center}
\includegraphics[width=\linewidth]{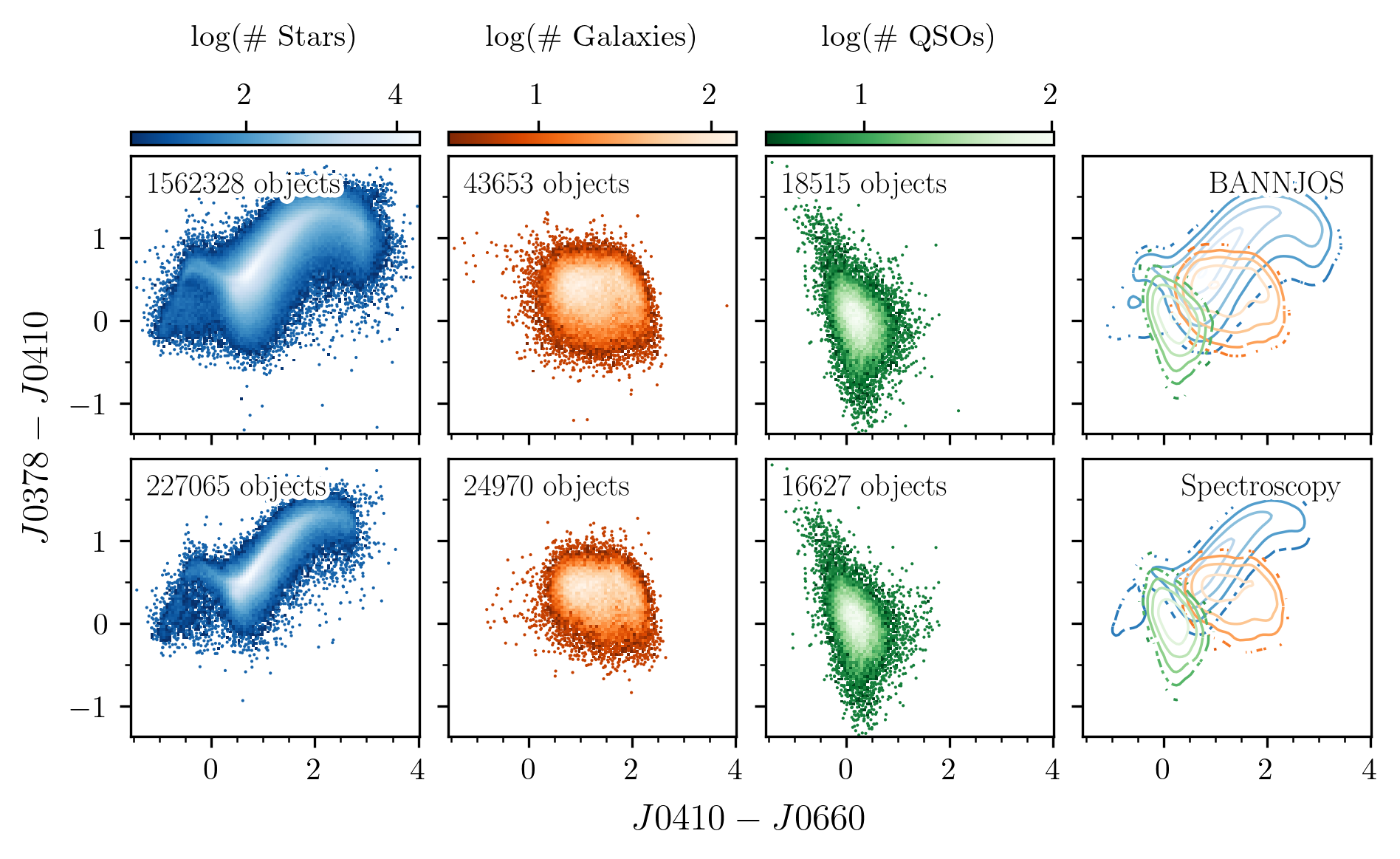}
\caption{Color-color diagrams for the three available classes. Stars, Galaxies, and QSOs are shown in blue, orange, and green respectively, with their distributions in the left, middle-left, and middle-right columns. The color shading corresponds to object density on a logarithmic scale, computed separately for each class. The right column shows the concentration of objects for each class using correspondingly colored contours. The top row shows \bannjos{} classification for the entire J-PLUS catalog based on the $2\sigma$ selection criterion. The bottom row shows the spectroscopic classification in the training sample. For clarity, we restricted the sample to sources with photometric uncertainties and reddening below 0.25 mag, and no photometric flags: 1,624,496 sources predicted by \bannjos{} (top panels) and 268,662 in the spectroscopic sample (bottom panels). The distribution of sources is very similar, indicating that \bannjos{} is effectively recovering the object classes even though colors were not used during model training.}
\label{fig:color_color}
\end{center}
\end{figure*}

In this section, we present color-color diagrams for all sources within the J-PLUS dataset that meet the $2\sigma$ selection criteria. It is important to note that colors were not directly employed in the model's training process. Consequently, observable color distinctions between different types of sources would indicate that the model has autonomously learned the relation between the J-PLUS photometric bands for each category. Thus, these diagrams act as additional validation of the classification. Ideally, a proficiently trained model would accurately identify the specific loci associated with each class; as a result, each class should occupy distinct areas within the color-color diagrams.

The top panels of Figure~\ref{fig:color_color} display the $J0378 - J0410$ versus $J0410 - J0660$ color-color diagram\footnote{Colors were chosen randomly between those showing a more obvious separation between classes, for illustrative purposes.} for the three classes selected with \bannjos{} using the $2\sigma$ criterion. The distribution of objects from the training list, which are spectroscopically confirmed, is shown in the bottom panels. For clarity, we limited both samples to only sources with photometric uncertainties and reddening below 0.25 mag. We also excluded sources with non-zero photometric flags, which would indicate poor measurements. Up to 1,624,496 sources predicted by \bannjos{} (top panels) and 268,662 in the spectroscopic sample (bottom panels) met these criteria. \bannjos{} appears to perform well in separating the three classes: stars, galaxies, and QSOs occupy their expected regions. Although some overlap is anticipated, the highest concentration for each individual class differs (rightmost panels). No significant trends in the isodensity contours suggest major misclassification from one species to another. The consistency of the locus of the three different classes, classified by \bannjos{} and by spectroscopy, indicates that our classifier can provide reasonably clean samples, comparable even to those obtained with spectroscopy.

Due to the imposed photometric quality cuts, Figure~\ref{fig:color_color} only includes sources with relatively high S/N. In fact, results using the $1\sigma$ and the $\max[PC_{\text{class}}(50)]$ criteria are nearly identical for this subset, as the limiting factor is the quality of the photometry. Relaxing the criteria for maximum allowed photometric error and reddening results in a more blurred locus for all three cases, yet consistent between the spectroscopic and photometric samples. Even without any quality selection in the photometry, the results remain consistent between \bannjos{} and the spectroscopic sample, though loci become much more blurred for \bannjos{} objects, which include very low S/N sources. In this scenario, the number of sources classified by \bannjos{} varies according to Table~\ref{tab:Final_counts}.

\section{Caveats}\label{sec:Caveats}

\subsection{Peculiar Tiles}\label{sec:Bad_Tiles}

During a thorough analysis of our results, we noticed that certain Tiles produced number counts far from expectations. We checked them in detail and we provide the list here for the shake of the user:

\begin{itemize}
    \item{Tiles 95100, 95118, 95126, 95136, 95162, 95175, 95186, 95197}. The number counts for stars are lower than expected at $r < 17$ mag. These Tiles have a factor of $10$ larger exposure time in the $r$ band than the rest of the Tiles, producing saturated stars to fainter magnitudes. 
    \item{Tile 98155}. An excess of galaxies and a lack of stars are observed at $r > 18$ mag. We noticed a relatively high photometric noise in the all the narrow bands that could be affecting the prediction.
    \item{Tile 102263}. An excess of both QSOs at all magnitudes and galaxies at $r > 18$ mag is observed. This pointing is dominated by M33, and the detection of structures of the galaxy as individual sources produces the observed overestimation.
    \item{Tile 103605}. Idem, but for M31.
    \item{Tiles 103874, 103884, 103908, 103919, 103930}. A lack of galaxies and QSOs at $r > 16$ mag is observed. These Tiles have the lower galactic latitude in the survey and the larger density of stars. We found that the detection from \texttt{SExtractor} is missing a large amount of faint sources and performed a deficient deblending in several cases.
\end{itemize}

We recommend using caution when utilizing data from these Tiles and leveraging the information from all other available classifiers.

\subsection{Multiple Entries}\label{sec:Multiple_entries}

We added data from other all-sky surveys into our data sets, including astrometry and photometry from the \Gaia{} third data release and the \Catwise{} catalog. The match between catalogs is offered along other scientific tables at J-PLUS data portal and is based on the sky positions of the sources, with a maximum match radius defined by the standard PSF of the surveys. Due to varying resolving powers of the surveys and atmospheric effects in J-PLUS, multiple matches are sometimes possible, especially for \Gaia{}, where the standard match radius of 1.5 arcsec allows more than one \Gaia{} source to be matched to a single J-PLUS source. Similar issues arise with \Catwise, whose wider PSF can result in multiple J-PLUS objects being matched to a single source. This leads to some objects having multiple entries in the training and output catalogs, potentially resulting in different classifications as the feature vector, $\mathbf{x}$, varies between entries (see Appendix~\ref{Appendix:Data_query}).

To be consistent with the J-PLUS archive and how cross-matches with external catalogs are provided, we decided to retain all possible classifications for any given object. This resulted in a list of  47,794,839 objects, 330,961 more than the original list. For some of these objects, the classification is consistent between entries. However, many vary depending on the matched source. We view this as beneficial: two different objects, blended by the J-PLUS PSF, can potentially be distinguished in such cases. Alternatively, users can select the most robust classification or simply exclude objects with multiple classifications.

\subsection{Robustness of Results}\label{sec:Robustness}

We meticulously selected, configured, and trained \bannjos{} to minimize potential biases and errors in its predictions. Despite our extensive cross-validation tests affirming its accuracy, machine learning models are limited by the quality of their target data, $y_{\mathrm{true}}$, and the features used for training, $\mathbf{x}$. Observational errors in these quantities can degrade the model's predictive performance. Additionally, systematic errors likely affecting the training data can lead to propagated biases in the final predictions, $y_{\mathrm{pred}}$.

Extensive testing was conducted to identify and analyze systematics in $y_{\mathrm{true}}$ within our training list. By construction, this list shows a large variety in terms of depth and color coverage. However, we found data to be generally consistent and of high quality. We manually reviewed the predictions from our best candidate models, which provided results qualitatively consistent between them, with our training lists, and previous studies. However, we cannot rule out the possibility of incorrect classifications by our method for given particular objects. Consequently, we advise users to exercise caution and utilize the covariance matrices and various percentiles to ensure sample purity, especially at low S/N, where the training and target samples differ the most.

\section{Conclusions}\label{sec:Conclusions}

In this study, we introduced \bannjos{}, the Bayesian Artificial Neural Network for the Javalambre Observatory Surveys. As a versatile machine learning pipeline, \bannjos{} leverages advanced deep Bayesian Neural Networks for regression tasks, proving itself as a comprehensive tool for analyzing and predicting a broad spectrum of numerical data. Specifically, we harnessed \bannjos{}'s capabilities to systematically classify an extensive dataset from the J-PLUS survey, categorizing an ensemble of 47.4 million astronomical objects into stars, QSOs, and galaxies. This classification is not merely categorical; \bannjos{} furnished detailed probability distribution functions (PDFs) for each source across the three categories, thus facilitating a nuanced selection process based on probabilistic confidence, inter-class correlations, and more.

The training of \bannjos{} was meticulously conducted using a list of approximately 1.2 million objects including approximately 430,000 galaxies (35\%), 120,000 QSOs (9\%), and 680,000 stars (55\%), all with reliable classifications from \SDSS{} DR18, \LAMOST{} DR9, \DESI{} EDR, or \Gaia{} DR3. We employed data augmentation techniques to balance the training sample, and incorporated information from 445 variables to the final training list, including photometric information in the 12 filters of J-PLUS across 8 different apertures, morphology variables, masking flags, image quality variables, etc., alongside infrared photometric information from \Catwise{} and astrometry from \Gaia. 

The results were validated using a test sample of approximately $1.4 \times 10^5$ sources. \bannjos{} demonstrated exceptional performance, with ROC AUC values greater than 0.99 for all six possible class combinations up to $r = 21.9$ mag, which corresponds to the 50\% completeness level in J-PLUS for compact sources. \bannjos{} also performs well at fainter magnitudes, correctly classifying 90\%, 81\%, and 87\% of galaxies, QSOs, and stars, respectively, at $21.5 < r \leq 22.5$ mag. The total accumulated error falls below 2\% for sources with $r \leq 22.5$ mag, and about 0.9\% at $r \leq 20.5$ mag. \bannjos{} significantly outperforms the three currently available classifiers in J-PLUS, \texttt{CLASS\_STAR}, \texttt{sglc\_prob\_star}, and the classification presented in \citet{vonmarttens24} with relative classification errors between 8 and 4 times smaller at $r \sim 22.5$ mag. Extensive tests revealed no significant biases or spatial trends. 

We have used \bannjos{} to classify all J-PLUS sources into around 20 million galaxies, 1 million QSOs, and 26 million stars. The resulting classification is consistent in number counts with results from previous works and model predictions. This consistency extends to magnitudes up to $r \sim 20.5$ mag for the three classes using the simplest classification criterion. However, results significantly improve when applying more restrictive criteria, aligning in number counts with model predictions up to $r \sim 21.5$ mag for the three species.

The full PDF provided by \bannjos{} enables J-PLUS users to refine their object selection. We have demonstrated its potential by selecting objects using three different criteria, allowing for the creation of very pure samples, even at faint magnitudes. Utilizing the covariance matrix allows for even finer selection of sources, enabling users to identify species not considered in the original classification, such as partially resolved active galaxies.

As general-purpose regressor, \bannjos{} can be utilized in a wide variety of scientific cases. For example to derive stellar chemical abundances, or photometric redshifts for galaxies or QSOs. Its potential will be further explored with the upcoming J-PAS survey, where the information from an astonishing amount of 56 colors will allow \bannjos{} to investigate the nature of each source in great detail.

The value-added catalog with the classification and whose content is described in Appendix~\ref{Appendix:Output} is accessible at the the J-PLUS archive through CEFCA's catalogues portal\footnote{\href{https://archive.cefca.es/catalogues/}{https://archive.cefca.es/catalogues/}}. This table, called \code{ClassBANNJOS}, has all the necessary information to reconstruct the full PDF, enabling astronomers to easily select sources based on their criteria and to resample \bannjos' predictions at their convenience. \bannjos{} and the code to decompress the PDF are publicly available at \url{https://github.com/AndresdPM/BANNJOS}. 

\begin{acknowledgements}
Based on observations made with the JAST80 telescope and T80Cam camera for the J-PLUS project at the Observatorio Astrof\'{\i}sico de Javalambre (OAJ), in Teruel, owned, managed, and operated by the Centro de Estudios de F\'{\i}sica del  Cosmos de Arag\'on (CEFCA). We acknowledge the OAJ Data Processing and Archiving Unit (UPAD) for reducing and calibrating the OAJ data used in this work. Funding for OAJ, UPAD, and CEFCA has been provided by the Governments of Spain and Arag\'on through the Fondo de Inversiones de Teruel and their general budgets; the Aragonese Government through the Research Groups E96, E103, E16\_17R, E16\_20R and E16\_23R; the Spanish Ministry of Science and Innovation (MCIN/AEI/10.13039/501100011033 y FEDER, Una manera de hacer Europa) with grants PID2021-124918NB-C41, PID2021-124918NB-C42, PID2021-124918NA-C43, and PID2021-124918NB-C44; the Spanish Ministry of Science, Innovation and Universities (MCIU/AEI/FEDER, UE) with grant PGC2018-097585-B-C21; the Spanish Ministry of Economy and Competitiveness (MINECO) under AYA2015-66211-C2-1-P, AYA2015-66211-C2-2, AYA2012-30789, and ICTS-2009-14; and European FEDER funding (FCDD10-4E-867, FCDD13-4E-2685). The Brazilian agencies FINEP, FAPESP, and the National Observatory of Brazil have also contributed to this project. This work has made use of data from the European Space Agency (ESA) mission \Gaia{} (\url{https://www.cosmos.esa.int/gaia}), processed by the \Gaia{} Data Processing and Analysis Consortium (DPAC, \url{https://www.cosmos.esa.int/web/gaia/dpac/consortium}). Funding for the DPAC has been provided by national institutions, in particular the institutions participating in the \Gaia{} Multilateral Agreement. F.J.E and P.C. acknowledge financial support from MCIN/AEI/10.13039/501100011033 through grant PID2020-112949GB-I00. C.H.M. also acknowledges the support of the Spanish Ministry of Science and Innovation via project grant PID2021-126616NB-I00, RvM is supported by Funda\c{c}\~ao de Amparo \`a Pesquisa do Estado da Bahia (FAPESB) grant TO APP0039/2023. VM thanks CNPq (Brazil) and FAPES (Brazil) for partial financial support. MQ is supported by the Brazilian research agencies FAPERJ and CNPq. A. del Pino acknowledges the financial support from the European Union - NextGenerationEU and the Spanish Ministry of Science and Innovation through the Recovery and Resilience Facility project ICTS-MRR-2021-03-CEFCA and the project PID2021-124918NB-C41. A. del Pino also thanks Dr. Bertran de Lis for her support and help during the realization of this project. 

{\it Software:} 
{\code{numpy} \citep{numpy20}, 
\code{scipy} \citep{scipy20},
\code{matplotlib} \citep{matplotib07}, 
\code{astropy} \citep{astropy13, astropy18, astropy22},
\code{tensorflow} \citep{tensorflow15}
}
\end{acknowledgements}

\bibliographystyle{aa}
\bibliography{biblio.bib}

\begin{appendix}

\section{Training Data Query and External Catalogs}\label{Appendix:Data_query}

We have enriched our training dataset with information from other all-sky surveys, specifically incorporating astrometry and photometry from the \Gaia{} third data release and the \Catwise{} catalog. For \Gaia{}, we utilized the standard match radius of $1^{\prime\prime}.5$ available in the J-PLUS archive, which typically provides a single match per J-PLUS object. However, for \Catwise{}, employing the standard matching radius of $5^{\prime\prime}$ (nearly equivalent to two ALLWISE pixels) resulted in approximately 20\% of J-PLUS sources being matched with multiple \Catwise{} counterparts. After a detailed examination, we determined that the majority, if not all, of these cases were spurious detections associated with local intensity maxima around extended sources, primarily galaxies. The cumulative distribution of angular distances between J-PLUS and \Catwise{} sources indicates that 99.9\% of matches occur within 1.5 arcseconds. Consequently, we opted to include only those \Catwise{} sources that are located within an angular distance of 1.5 arcseconds from a J-PLUS source, while retaining all matches that have smaller separations.

It is important to note that, by adhering to this criterion, some objects may appear more than once in the catalog. This situation could, in theory, lead to multiple different outcomes from \bannjos{}. However, for the sake of completeness and consistency with the matching criteria used for other external catalogs in the J-PLUS archive, we decided to retain all such matches.

For each Tile, we performed exactly the same query. For example, our query for the Tile 103930 was:

\code{SELECT\ FLambda.*,\ MagAB.MU\_MAX,\ MagAB.APER3\_WORSTPSF,\ FNu.FLUX\_AUTO,\ FNu.FLUX\_APER\_0\_8,\ FNu.FLUX\_APER\_1\_0,\ FNu.FLUX\_APER\_1\_2,\ FNu.FLUX\_APER\_1\_5\,\ FNu.FLUX\_APER\_2\_0,\ FNu.FLUX\_APER\_3\_0,\ FNu.FLUX\_APER\_4\_0,\ FNu.FLUX\_APER\_6\_0,\ MagAB.MAG\_AUTO,\ MagAB.MAG\_APER\_0\_8,\ MagAB.MAG\_APER\_1\_0,\ MagAB.MAG\_APER\_1\_2,\ MagAB.MAG\_APER\_1\_5,\ MagAB.MAG\_APER\_2\_0,\ MagAB.MAG\_APER\_3\_0,\ MagAB.MAG\_APER\_4\_0,\ MagAB.MAG\_APER\_6\_0,\ FNu.FLUX\_RELERR\_AUTO,\ FNu.FLUX\_RELERR\_APER\_0\_8,\ FNu.FLUX\_RELERR\_APER\_1\_0,\ FNu.FLUX\_RELERR\_APER\_1\_2,\ FNu.FLUX\_RELERR\_APER\_1\_5,\ FNu.FLUX\_RELERR\_APER\_2\_0,\ FNu.FLUX\_RELERR\_APER\_3\_0,\ FNu.FLUX\_RELERR\_APER\_4\_0,\ FNu.FLUX\_RELERR\_APER\_6\_0,\ MagAB.MAG\_ERR\_AUTO,\ MagAB.MAG\_ERR\_APER\_0\_8,\ MagAB.MAG\_ERR\_APER\_1\_0,\ MagAB.MAG\_ERR\_APER\_1\_2,\ MagAB.MAG\_ERR\_APER\_1\_5,\ MagAB.MAG\_ERR\_APER\_2\_0,\ MagAB.MAG\_ERR\_APER\_3\_0,\ MagAB.MAG\_ERR\_APER\_4\_0,\ MagAB.MAG\_ERR\_APER\_6\_0,\ MWEx.ax,\ MWEx.ax\_err,\ MWEx.ebv,\ MWEx.ebv\_err,\ gaia.angDist,\ gaia.pmra\ as\ pmra\_g,\ gaia.pmde,\ gaia.plx,\ gaia.ruwe,\ gaia.fg,\ gaia.fbp,\ gaia.frp,\ gaia.e\_pmra,\ gaia.e\_pmde,\ gaia.e\_plx,\ gaia.e\_fg,\ gaia.e\_fbp,\ gaia.e\_frp,\ allwise.angDist,\ allwise.Jmag,\ allwise.Hmag,\ allwise.Kmag,\ allwise.e\_Jmag,\ allwise.e\_Hmag,\ allwise.e\_Kmag,\ \ catwise.angDist,\ catwise.pmRA,\ catwise.e\_pmRA,\ catwise.pmDE,\ catwise.e\_pmDE,\ catwise.W1mproPM,\ catwise.e\_W1mproPM,\ catwise.W2mproPM,\ catwise.e\_W2mproPM FROM\ jplus.FLambdaDualObj\ AS\ FLambda\ LEFT\ JOIN\ jplus.MagABDualObj\ AS\ MagAB\ ON\ ((FLambda.NUMBER\ =\ MagAB.NUMBER)\ AND\ (FLambda.TILE\_ID\ =\ MagAB.TILE\_ID))\ LEFT\ JOIN\ jplus.FNuDualObj\ AS\ FNu\ ON\ ((FLambda.NUMBER\ =\ FNu.NUMBER)\ AND\ (FLambda.TILE\_ID\ =\ FNu.TILE\_ID))\ LEFT\ JOIN\ jplus.MWExtinction\ AS\ MWEx\ ON\ ((FLambda.NUMBER\ =\ MWEx.NUMBER)\ AND\ (FLambda.TILE\_ID\ =\ MWEx.TILE\_ID))\ LEFT\ JOIN\ jplus.xmatch\_gaia\_dr3\ AS\ gaia\ ON\ ((FLambda.NUMBER\ =\ gaia.NUMBER)\ AND\ (FLambda.TILE\_ID\ =\ gaia.TILE\_ID))\ LEFT\ JOIN\ jplus.xmatch\_allwise\ AS\ allwise\ ON\ ((FLambda.NUMBER\ =\ allwise.NUMBER)\ AND\ (FLambda.TILE\_ID\ =\ allwise.TILE\_ID))\ LEFT\ JOIN\ jplus.xmatch\_catwise2020\ AS\ catwise\ ON\ ((FLambda.NUMBER\ =\ catwise.NUMBER)\ AND\ (FLambda.TILE\_ID\ =\ catwise.TILE\_ID))\ WHERE\ ((catwise.angDist\ <=\ 1.5)\ AND\ (FLambda.TILE\_ID\ =\ 103930))}

\section{Model Selection and Hyper-Parameter Tuning}\label{Appendix:Model_Selection}

\begin{figure*}
\begin{center}
\includegraphics[width=\linewidth]{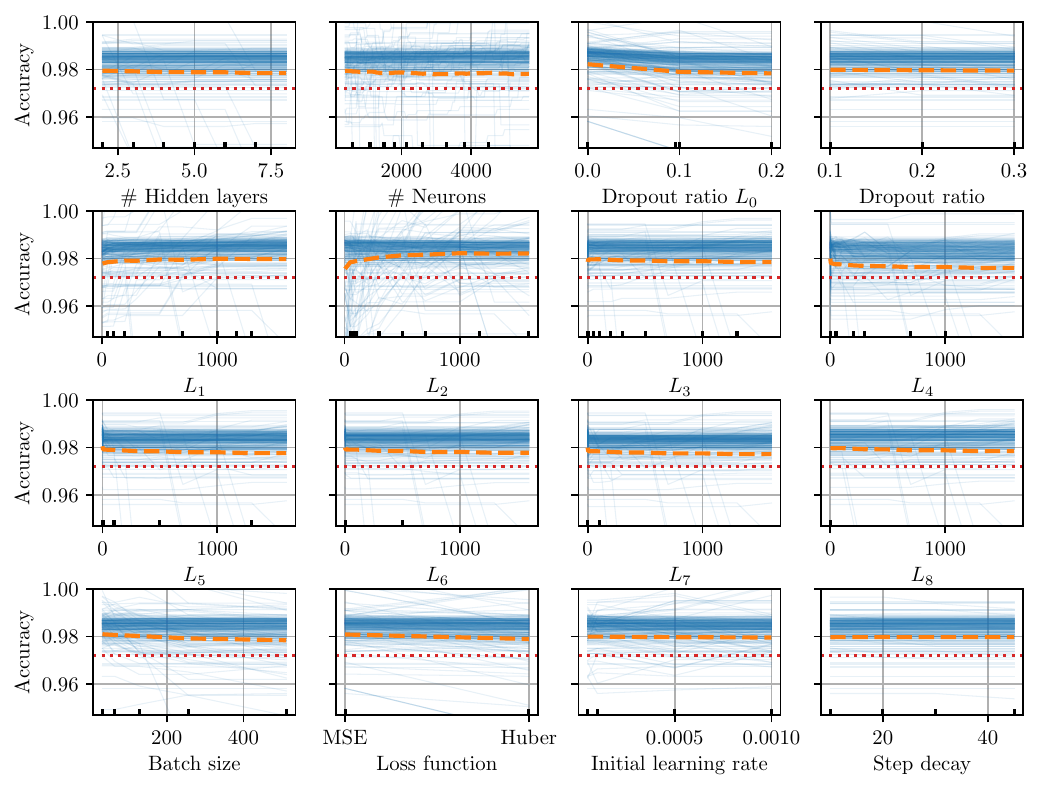}
\caption{Individual Conditional Expectation curves for the model accuracy. Each blue line represents the accuracy of the model depending on the chosen value for each hyperparameter. The orange dashed line indicates the mean accuracy value, approximately 0.98, which is below the median accuracy value, closer to 0.99. The red dashed line represents the performance of a Random Forest Regressor with 500 trees. Most hyperparameters have minimal impact on model performance, except for the Dropout at $L_0$, and the number of neurons at levels $L_{1-4}$.}
\label{fig:ICE_hyper}
\end{center}
\end{figure*}

Figure~\ref{fig:ICE_hyper} displays Individual Conditional Expectation (ICE) curves illustrating the model's accuracy as influenced by each hyperparameter. We note variations in performance depending on the hyperparameters employed, observing generally better outcomes with deeper architectures and notably the number of neurons in the first four hidden layers ($L_{1-4}$). Optimal accuracy was achieved with models containing a large number of neurons in $L_1$ and $L_2$ (at least 200) followed by additional hidden layers with fewer neurons. Our tests revealed correlations between certain hyperparameters, such as simultaneous increases in $L_1$ and $L_2$ or $L_2$ and $L_5$, leading to improved results. Conversely, some degree of anticorrelation between $L_3$ and $L_{4-5}$, and between $L_4$ and $L_5$, suggests a balance between the number of layers and free parameters for maintaining accuracy. The neuron count in layers $L_5$ to $L_8$ had minimal impact on accuracy, though a slight decline was observed with higher neuron counts ($\gtrsim 500$). Similarly, increasing the dropout ratio at $L_0$ marginally reduced model accuracy. Other parameters had less significant effects on performance, with the dropout BANN maintaining consistent average performance for models featuring three or more hidden layers. The best hyperparameters were identified by a Histogram-based Gradient Boosting Regression Tree, and are detailed in Section~\ref{sec:Model_Selection}.

\section{Data Compression}\label{Appendix:Data_compression}

\begin{figure}
\begin{center}
\includegraphics[width=\linewidth]{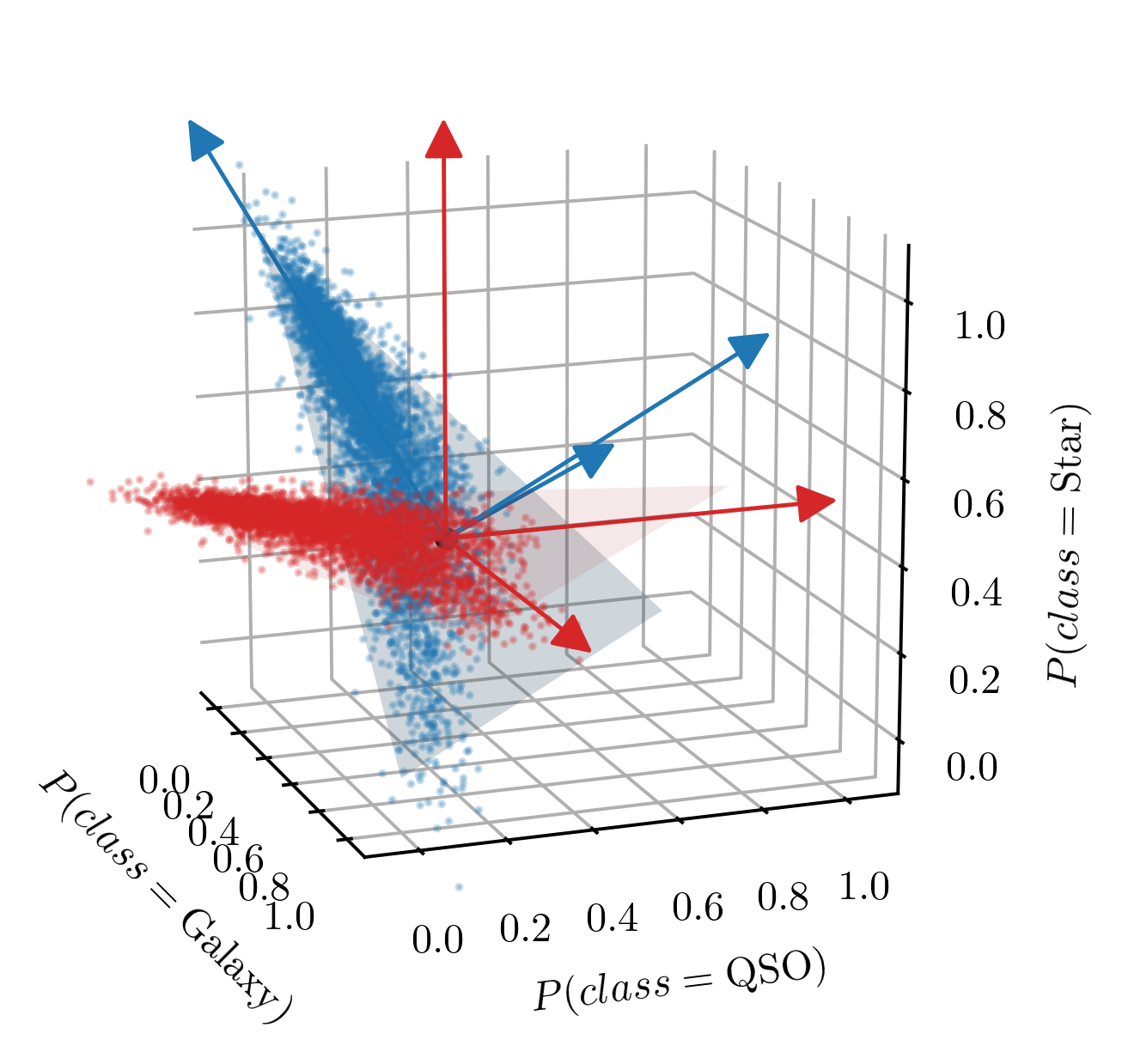}
\caption{Sampling from the posterior of the model for {\tt Tile Id} = 94217 and {\tt Number} = 10482, a source with low signal-to-noise ratio. Each blue point represents a sample from the BANN ($N = 5000$). The most probable class is 'star', though \bannjos{} also assigns some probability to the 'galaxy' class. The points distribute on the plane $P(\text{class} = \text{Galaxy}) + P(\text{class} = \text{QSO}) + P(\text{class} = \text{Star}) = 1$, as expected. Blue arrows represent the eigenvectors of the 3-dimensional space defined by the probabilities of belonging to each of the three classes. A rotation (Equation~\ref{eq:rotation}) applied to the blue points allows for dimension reduction, transforming them into the red points on the horizontal plane. The red arrows denote the eigenvectors of this plane.}
\label{fig:3D_Distro}
\end{center}
\end{figure}

Figure~\ref{fig:3D_Distro} illustrates the results for a source with ambiguous classification by the model ({\tt Tile Id} = 94217, {\tt Number} = 10482) with $N = 5000$. As anticipated, the points reside on the plane $P(\text{class} = \text{Galaxy}) + P(\text{class} = \text{QSO}) + P(\text{class} = \text{Star}) = 1$ with minimal dispersion, indicating the model's correct learning of the correlations between the three probabilities without explicit training.

Although retaining all points permits full PDF reconstruction for the object, extensive sampling with large $N$ entails high computational and storage costs. To mitigate this, we experimented with smaller $N$ values, identifying a minimal consistent value, $N=300$. However, this still results in substantial amount of data. To further reduce data volume, we exploit the points' planar distribution by applying a rotation:

\begin{equation}
    R = 
    \begin{pmatrix}
    (\sqrt{3}+3)/6         & -\sqrt{(2-\sqrt{3})/6} & -1/\sqrt{3} \\
    -\sqrt{(2-\sqrt{3})/6} & (\sqrt{3}+3)/6         & -1/\sqrt{3} \\
    1/\sqrt{3}             & 1/\sqrt{3}             & 1/\sqrt{3}
    \end{pmatrix}
\end{equation}\label{eq:rotation},

which projects the $N \times 3$ points onto the plane defined by $P(\text{x} = \text{Galaxy}) + P(\text{x} = \text{QSO}) + P(\text{x} = \text{Star}) = 1$. In this projection, the vertical coordinate, perpendicular to the plane, is negligible and can be discarded. This transformation leverages the interrelated PDFs to shift from three to two dimensions, effectively reducing storage needs to $N \times 2$. Further compression is achieved by fitting a Gaussian Mixture Model, GMM, with three components to the $N \times 2$ points, parameterizing the entire PDF with the means, covariances, and weights of the three Gaussian distributions—totalling 18 parameters after reducing the covariance matrix elements.

This compression approach was tested with sources exhibiting one, two, and three components in their predicted PDFs by creating an additional validation sample through $N = 5000$ BANN posterior samples ($\text{PDF}_{\text{hi}}$). We then fitted GMMs to both high and low-resolution datasets, reconstructing 3D PDFs by sampling each fitted GMM 5000 times. For each test sample object, we thus compared four different PDFs: $\text{PDF}_{\text{hi}}$, $\text{PDF}_{\text{lo}}$, $\text{GMM}_{\text{hi}}$, and $\text{GMM}_{\text{lo}}$, adopting the class with the highest median value from each PDF as the predicted classification.

The vast majority of sources showed consistent classifications across $\text{PDF}_{\text{hi}}$, $\text{PDF}_{\text{lo}}$, and their corresponding GMMs. Furthermore, $\text{GMM}_{\text{lo}}$ accurately replicated the original $\text{PDF}_{\text{hi}}$ in all visually inspected cases. However, a small fraction of cases exhibited classification shifts due to reduced sample size and the GMM compression, affecting less than 0.003\% of cases (4 out of 136570), typically those with low S/N and large classification uncertainties. These discrepancies were mitigated by applying quality cuts. For the remainder, \bannjos{} yielded consistent results between high-quality and standard samples and between original and reconstructed PDFs. Considering the significant benefits and minimal disadvantages, we proceeded with $N = 300$ and GMM compression.

\section{Selecting pure samples}\label{Appendix:Selecting_pure_samples}

In this Appendix, we provide some examples of Astronomical Data Query Language (ADQL) queries that yield purer object selections. Using the percentiles of the PDF, users can specify precisely the type of objects they wish to select. For example, the following query selects stars at the $1\sigma$ confidence level and provides their sky coordinates and $r$ magnitudes:

\code{SELECT\ MagAB.ALPHA\_J2000,\ MagAB.DELTA\_J2000,\ MagAB.MAG\_PSFCOR[1],\ BANNJOS.*\ FROM\ jplus.MagABDualObj\ AS\ MagAB\ LEFT\ JOIN\ jplus.ClassBANNJOS\ AS\ BANNJOS\ ON\ ((BANNJOS.NUMBER\ =\ MagAB.NUMBER)\ AND\ (BANNJOS.TILE\_ID\ =\ MagAB.TILE\_ID))\ WHERE\ (BANNJOS.CLASS\_STAR\_prob\_pc16\ >=\ 0.333)}

Here, we require the 16th percentile to be above $\frac{1}{3}$ of the probability, which is equivalent to a random probability in a classification with three possible classes. Since the probability at which a certain percentile is fulfilled does not need to be complementary between classes, this query also selects objects whose probability of being a galaxy or a QSO is compatible within one sigma with $\frac{1}{3}$. To be even more restrictive when selecting stars, we can select only sources with the 84th percentiles below $\frac{1}{3}$:

\code{SELECT\ MagAB.ALPHA\_J2000,\ MagAB.DELTA\_J2000,\ MagAB.MAG\_PSFCOR[1],\ BANNJOS.*\ FROM\ jplus.MagABDualObj\ AS\ MagAB\ LEFT\ JOIN\ jplus.ClassBANNJOS\ AS\ BANNJOS\ ON\ ((BANNJOS.NUMBER\ =\ MagAB.NUMBER)\ AND\ (BANNJOS.TILE\_ID\ =\ MagAB.TILE\_ID))\ WHERE\ ((BANNJOS.CLASS\_STAR\_prob\_pc16\ >=\ 0.333)\ AND\ (BANNJOS.CLASS\_QSO\_prob\_pc84\ <\ 0.333)\ AND\ (BANNJOS.CLASS\_GALAXY\_prob\_pc84\ <\ 0.333))}

The same criteria can be used to select very probable QSOs at the $2\sigma$ confidence level. Furthermore, we can use the additional information from the correlation between the probabilities to select objects with a negative correlation between being a star and a QSO. Requiring negative correlations between the classes' probabilities can be interpreted as requiring that there is no confusion between the classes, i.e., it is either one or the other.

\code{SELECT\ MagAB.ALPHA\_J2000,\ MagAB.DELTA\_J2000,\ MagAB.MAG\_PSFCOR[1],\ BANNJOS.*\ FROM\ jplus.MagABDualObj\ AS\ MagAB\ LEFT\ JOIN\ jplus.ClassBANNJOS\ AS\ BANNJOS\ ON\ ((BANNJOS.NUMBER\ =\ MagAB.NUMBER)\ AND\ (BANNJOS.TILE\_ID\ =\ MagAB.TILE\_ID))\ WHERE\ ((BANNJOS.CLASS\_QSO\_prob\_pc02\ >=\ 0.333)\ AND\ (BANNJOS.CLASS\_STAR\_prob\_pc98\ <\ 0.333)\ AND\ (BANNJOS.CLASS\_GALAXY\_prob\_pc98\ <\ 0.333)\ AND\ (BANNJOS.CLASS\_QSO\_CLASS\_STAR\_prob\_corr\ <=\ 0))}

This query will return a sample of 404,070 high-confidence QSOs in J-PLUS. We tested this sample by cross-matching its objects with those from our test sample, obtaining a total of 11,230 common sources never seen by \bannjos{} during training. From these sources, 134 are spectroscopically classified as galaxies, and 18 as stars. This represents approximately $1.19\%$ and $0.16\%$ contamination from these other species, respectively. It is important to point out, however, that we did not explicitly reject sources with positive correlation between the galaxy and QSO classes, since we could also be interested in partially resolved active galaxies. Requiring \code{BANNJOS.CLASS\_GALAXY\_CLASS\_QSO\_prob\_corr\ <=\ 0} would produce an even cleaner sample by reducing the contamination from these types of sources.

We cross-matched this list of high-confidence QSOs from our training list, obtaining up to 290,794 new QSO candidates not previously cataloged either by \SDSS, \LAMOST, \DESI, or \Gaia. If we relax the conditions to $1\sigma$, which still produces very pure samples, this number increases to 388,976 relatively high-confidence QSOs without spectroscopic confirmation that could be used for spectroscopic follow-up.

\section{Performance comparison between classifiers}\label{Appendix:vM24_star}

\begin{figure*}
\begin{center}
\includegraphics[width=\linewidth]{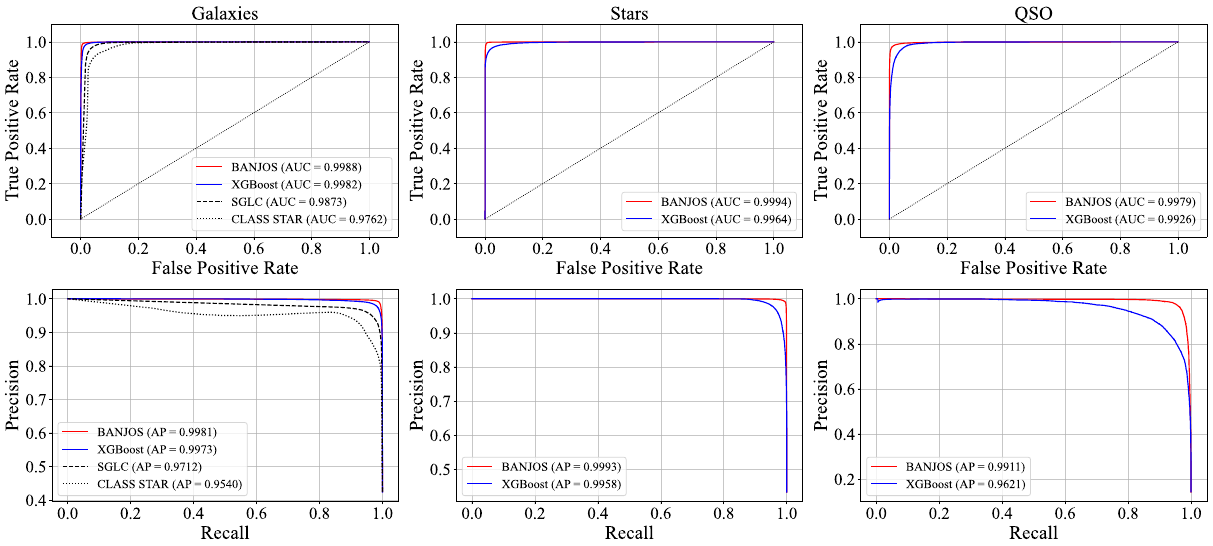}
\caption{{\it Top:} ROC curves for the three classes using the four available classifiers. \code{sglc\_prob\_star} and \code{CLASS\_STAR} are only shown for extended sources. {\it Bottom:} Corresponding PR curves.}
\label{fig:ROC_comparison}
\end{center}
\end{figure*}

\begin{figure*}
\begin{center}
\includegraphics[width=\linewidth]{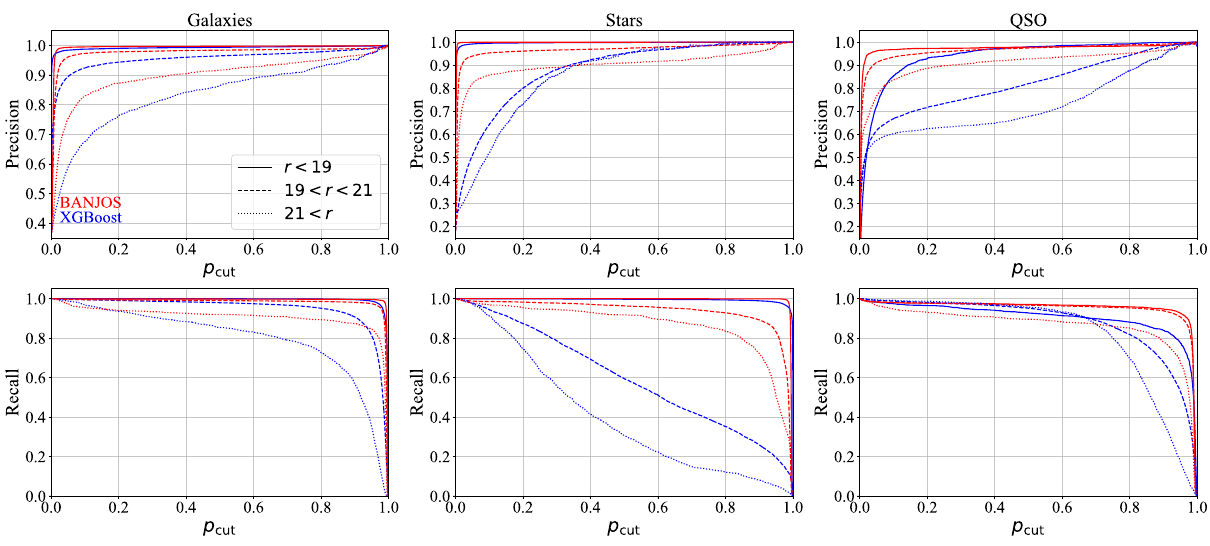}
\caption{Precision and Recall . {\it Top:} Precision curves for vM24 and \bannjos{} for the three classes in three different magnitude bins. The precision is computed as function of the probability used to select each class ($p_{cut}$). In the case of \bannjos{} we use the maximum of the median probability. {\it Bottom:} Corresponding Recall curves.}
\label{fig:Purity_completeness_comparison}
\end{center}
\end{figure*}

In this Appendix, we present a more general analysis of the comparison between the classifiers presented in Section~\ref{sec:Comparison_previous} using the test sample of sources not common to the training of \bannjos{} or the \code{XGBoost} algorithm used in vM24. Figure~\ref{fig:ROC_comparison} shows the ROC and the PR curves for the four analyzed classifiers. We assign the class predicted by \bannjos{} as the one with the highest median probability, $\max[PC_{\text{class}}(50)]$. For the rest of the classifiers, the class is selected as the one with the higher probability. Point-like sources have been separated into stars and QSOs only for vM24 and \bannjos{}. Figure~\ref{fig:Purity_completeness_comparison} shows the Precision and the Recall curves separately for the vM24 classification and \bannjos{} at different magnitude bins. As discussed in detail in Section~\ref{sec:Comparison_previous}, \bannjos{} is superior to all other classifiers for all the considered classes and across the entire magnitude range. It is also noticeable that the \code{XGBoost} algorithm used in vM24 is superior to \code{sglc\_prob\_star} and \code{CLASS\_STAR} for classifying extended sources, with \code{CLASS\_STAR} being the least reliable of all four classifiers.

\section{The \code{ClassBANNJOS} table}\label{Appendix:Output}

Table~\ref{tab:Output_structure} details the structure of the \code{ClassBANNJOS} table generated by \bannjos{}. For additional information on the significance of each field, refer to Section~\ref{sec:training_sampling} and Appendix~\ref{Appendix:Data_compression}. The table is accessible through J-PLUS data portal\footnote{\href{https://archive.cefca.es/catalogues/jplus-dr3}{https://archive.cefca.es/catalogues/jplus-dr3}}, named as \code{ClassBANNJOS}.

\begin{table*}
\caption{Structure of the \code{ClassBANNJOS} table.}
\label{tab:Output_structure}
\centering 
        \begin{tabular}{l c l}
        \hline\hline\rule{0pt}{3ex} 
        Name   & Type  & Description \\ 
        \hline
\texttt{TILE\_ID} &                               \texttt{int}   &  Identifier of the Tile image where the object was detected                                     \\
\texttt{NUMBER} &                                 \texttt{int}   &  Number identifier assigned by Sextractor for the object in the image                           \\
\texttt{CLASS\_GALAXY\_prob\_mean} &              \texttt{float} &  Mean probability of being a galaxy                                                             \\
\texttt{CLASS\_QSO\_prob\_mean} &                 \texttt{float} &  Mean probability of being a QSO                                                                \\
\texttt{CLASS\_STAR\_prob\_mean} &                \texttt{float} &  Mean probability of being a star                                                               \\
\texttt{CLASS\_GALAXY\_prob\_std} &               \texttt{float} &  MAD standard deviation of the probability of being a galaxy                                    \\
\texttt{CLASS\_QSO\_prob\_std} &                  \texttt{float} &  MAD standard deviation of the probability of being a QSO                                       \\
\texttt{CLASS\_STAR\_prob\_std} &                 \texttt{float} &  MAD standard deviation of the probability of being a star                                      \\
\texttt{CLASS\_GALAXY\_CLASS\_QSO\_prob\_corr} &  \texttt{float} &  Correlation between the probabilities of being a galaxy or a QSO                               \\
\texttt{CLASS\_GALAXY\_CLASS\_STAR\_prob\_corr} & \texttt{float} &  Correlation between the probabilities of being a galaxy or a star                              \\
\texttt{CLASS\_QSO\_CLASS\_STAR\_prob\_corr} &    \texttt{float} &  Correlation between the probabilities of being a QSO or a star                                 \\
\texttt{CLASS\_GALAXY\_prob\_pc02} &              \texttt{float} &  2.275th percentile of the probability of being a galaxy                                         \\
\texttt{CLASS\_QSO\_prob\_pc02} &                 \texttt{float} &  2.275th percentile of the probability of being a QSO                                            \\
\texttt{CLASS\_STAR\_prob\_pc02} &                \texttt{float} &  2.275th percentile of the probability of being a star                                           \\
\texttt{CLASS\_GALAXY\_prob\_pc16} &              \texttt{float} &  15.865th percentile of the probability of being a galaxy                                        \\
\texttt{CLASS\_QSO\_prob\_pc16} &                 \texttt{float} &  15.865th percentile of the probability of being a QSO                                           \\
\texttt{CLASS\_STAR\_prob\_pc16} &                \texttt{float} &  15.865th percentile of the probability of being a star                                          \\
\texttt{CLASS\_GALAXY\_prob\_pc50} &              \texttt{float} &  Median of the probability of being a galaxy                                                     \\
\texttt{CLASS\_QSO\_prob\_pc50} &                 \texttt{float} &  Median of the probability of being a QSO                                                       \\
\texttt{CLASS\_STAR\_prob\_pc50} &                \texttt{float} &  Median of the probability of being a star                                                      \\
\texttt{CLASS\_GALAXY\_prob\_pc84} &              \texttt{float} &  84.135th percentile of the probability of being a galaxy                                        \\
\texttt{CLASS\_QSO\_prob\_pc84} &                 \texttt{float} &  84.135th percentile of the probability of being a QSO                                           \\
\texttt{CLASS\_STAR\_prob\_pc84} &                \texttt{float} &  84.135th percentile of the probability of being a star                                          \\
\texttt{CLASS\_GALAXY\_prob\_pc98} &              \texttt{float} &  97.725th percentile of the probability of being a galaxy                                        \\
\texttt{CLASS\_QSO\_prob\_pc98} &                 \texttt{float} &  97.725th percentile of the probability of being a QSO                                           \\
\texttt{CLASS\_STAR\_prob\_pc98} &                \texttt{float} &  97.725th percentile of the probability of being a star                                          \\
\texttt{comp1\_cov\_11} &                         \texttt{float} &  2-dimensional covariance term 11 for Gaussian component 1                                      \\
\texttt{comp1\_cov\_12} &                         \texttt{float} &  2-dimensional covariance term 12 for Gaussian component 1                                      \\
\texttt{comp1\_cov\_22} &                         \texttt{float} &  2-dimensional covariance term 22 for Gaussian component 1                                      \\
\texttt{comp2\_cov\_11} &                         \texttt{float} &  2-dimensional covariance term 11 for Gaussian component 2                                      \\
\texttt{comp2\_cov\_12} &                         \texttt{float} &  2-dimensional covariance term 12 for Gaussian component 2                                      \\
\texttt{comp2\_cov\_22} &                         \texttt{float} &  2-dimensional covariance term 22 for Gaussian component 2                                      \\
\texttt{comp3\_cov\_11} &                         \texttt{float} &  2-dimensional covariance term 11 for Gaussian component 3                                      \\
\texttt{comp3\_cov\_12} &                         \texttt{float} &  2-dimensional covariance term 12 for Gaussian component 3                                      \\
\texttt{comp3\_cov\_22} &                         \texttt{float} &  2-dimensional covariance term 22 for Gaussian component 3                                      \\
\texttt{comp1\_mean\_1} &                         \texttt{float} &  2-dimensional mean term 1 for Gaussian component 1                                             \\
\texttt{comp1\_mean\_2} &                         \texttt{float} &  2-dimensional mean term 2 for Gaussian component 1                                             \\
\texttt{comp2\_mean\_1} &                         \texttt{float} &  2-dimensional mean term 1 for Gaussian component 2                                             \\
\texttt{comp2\_mean\_2} &                         \texttt{float} &  2-dimensional mean term 2 for Gaussian component 2                                             \\
\texttt{comp3\_mean\_1} &                         \texttt{float} &  2-dimensional mean term 1 for Gaussian component 3                                             \\
\texttt{comp3\_mean\_2} &                         \texttt{float} &  2-dimensional mean term 2 for Gaussian component 3                                             \\
\texttt{comp1\_weight} &                          \texttt{float} &  2-dimensional weight for Gaussian component 1                                                  \\
\texttt{comp2\_weight} &                          \texttt{float} &  2-dimensional weight for Gaussian component 2                                                  \\
\texttt{comp3\_weight} &                          \texttt{float} &  2-dimensional weight for Gaussian component 3                                                  \\
        \hline 
\end{tabular}
\end{table*}

\end{appendix}

\end{document}